\documentclass[journal]{IEEEtran}
\newcommand{\indicator}[1]{\mathbbm{1}_{#1}}

\usepackage{cite}
\usepackage{epsfig}
\usepackage{amsmath}
\usepackage{amssymb}
\usepackage{bbm}
\usepackage[caption=false]{subfig}
\usepackage{hyperref}

\begin{document}

\title{A highly optimized flow-correlation attack}
\author{Juan~A.~Elices,~\IEEEmembership{Student Member,~IEEE,}~and~Fernando~P\'erez-Gonz\'alez,~\IEEEmembership{Senior Member,~IEEE}
\thanks{Copyright (c) 2013 IEEE. Personal use of this material is permitted. However, permission to use this material for any other purposes must be obtained from the IEEE by sending a request to pubs-permissions@ieee.org.}
\thanks{J.A.Elices is with the Dept. of Electrical and Computer Engineering, University of New Mexico, Albuquerque, NM, 87131-0001 USA, e-mail: jelices@ece.unm.edu }
\thanks{Prof. F. P\'erez-Gonz\'alez is with the Signal Theory and Communications Department, University of Vigo, Vigo 36310, SPAIN, with the Galician Research and Development Center in Advanced Telecommunications (GRADIANT), Vigo 36310,
SPAIN, and with the Dept. of Electrical and Computer Engineering, University of New Mexico, Albuquerque, NM, 87131-0001 USA, e-mail: fperez@gts.uvigo.es}}

\maketitle

\begin{abstract}
Deciding that two network flows are essentially the same is an important problem in intrusion detection and in tracing anonymous connections. A stepping stone or an anonymity network may try to prevent flow correlation by adding chaff traffic, splitting the flow in several subflows or adding random delays.  A well-known attack for these types of systems is active watermarking. However, active watermarking systems can be detected and an attacker can modify the flow in such a way that the watermark is removed and can no longer be decoded. This leads to the two basic features of our scheme: a highly-optimized algorithm that achieves very good performance and a passive analysis that is undetectable. 

We propose a new passive analysis technique where detection is based on Neyman-Pearson lemma. We correlate the inter-packet delays (IPDs) from both flows. Then, we derive a modification to deal with stronger adversary models that add chaff traffic, split the flows or add random delays. We empirically validate the detectors with a simulator. Afterwards, we create a watermark-based version of our scheme to study the trade-off between performance and detectability. Then, we compare the results with other state-of-the-art traffic watermarking schemes in several scenarios concluding that our scheme outperforms the rest. Finally, we present results using an implementation of our method on live networks, showing that the conclusions can be extended to real-world scenarios. 

Our scheme needs only tens of packets under normal network interference and a few hundreds of packets when a number of countermeasures are taken.
\end{abstract}


\IEEEpeerreviewmaketitle

\section{Introduction}

Network attackers intentionally hide their identity to avoid prosecution. A widely-used way of achieving this anonymity is forwarding the traffic through a chain of compromised hosts called stepping stones~\cite{StHe:95}. Tracing back the chain to the source is a challenging problem due to the encrypted or even anonymized connections between stepping stones. Deciding that two flows are essentially the same can be applied to the mentioned problem as well as in many other contexts, such as tracing anonymous connections~\cite{SyTsReLa:01} or preventing congestion attacks on anonymous networks~\cite{HoKi:11}.

There are two general approaches for finding correlated flows: passive analysis and active watermarks. Passive analysis schemes are based on correlating some characteristics of the flows, such as packet timings or packet counts, without altering such flows~\cite{ZhPa:00,DoFlShPaCoStL:02, BlSoVe:04}.  On the other hand, active watermarks actively modify the flow by delaying individual packets. Active watermarks can be packet-based, embedding the watermark on individual
delays between packets~\cite{WaRe:03,HoKiBo2:09} or interval-based, embedding the watermark in some properties of the intervals~\cite{HoKi:11,PyPaReWaNi:12,WaChJa:07}.

Lately, most of the work has been focused on the design of new active watermarking techniques, as they are considered to be more efficient, obtaining lower error rates for flows of the same length. All of them are designed with the idea
of being undetectable, as detection can lead to the stepping stone or the anonymous network to modify the flow in such a way that it it can no longer be detected. In spite of this, detecting these watermarks has been shown to be feasible and not a hard task~\cite{LuZhZhPeLe:11, LiHo:12}. This allows an attacker, e.g. the stepping stone, anonymous network, etc., to easily modify the timing of the detected flow to prevent the correlation using techniques as chaff packets, flow splitting, merging flows or adding random delays.

Achieving a good performance for flow-correlation approaches is critical for two main reasons: first, to be able to deal with flow modifications and other countermeasures, and second, to ensure a minimum reliability, implying a small probability of false positives.
These two reasons lead to the necessity of extremely accurate techniques to correlate flows as provided by our scheme. Furthermore, we cannot rely on the length of the sequence, as in many kinds of stealthy attacks, the amount of traffic sent by the attackers or compromised bots is very small.

We propose a passive traffic analysis technique that outperforms any of the state-of-the-art traffic watermarking schemes. For instance, 21 packets separated at least 10 ms are enough to correlate two flows, one in Virginia, the other in California, correctly with probability $0.9861$ when the false positive probability is fixed to $10^{-5}$ and no countermeasures are exerted. The proposed method saves the inter-packet delays (IPDs) of the flow and uses a detector based on the likelihood ratio test (Neyman-Pearson lemma). 

As IPDs are not robust against the insertion and drop of packets, we develop a modification which is robust against chaff packets, repacketization, flow splitting, and attacks that add or remove packets from the flow. We also make it robust against random delays under a maximum delay constraint. 

The rest of the paper is organized as follows: Section \ref{sec:bg} reviews previous schemes for correlating flows and techniques for detecting active watermarks. In Section \ref{sec:sc} we introduce the notation that we follow and formally describe the model. In Section \ref{sc:det1} we construct our detector. Section \ref{sc:prf} validates its performance using a simulator. Section \ref{sc:det2} proposes a modification to ensure robustness against chaff traffic, flow splitting and constrained random delays. In Section \ref{sc:watvspa} we create an active watermark to study the trade-off between performance and detectability. In Section \ref{sc:com} we compare our passive scheme with existing algorithms in terms of error probability. Section \ref{sc:ri} shows the result of a real implementation. Finally, Section \ref{sc:con} summarizes our contribution. 

\section{Previous Work}\label{sec:bg}

\subsection{Passive analysis}

Zang and Paxson~\cite{ZhPa:00} proposed to correlate the traffic by measuring the time that both flows are in OFF (i.e., no transmission) state. They achieve a large confidence when connections are several minutes long (i.e., thousands of packets) but not so much reliability on short connections. They do not consider any alteration in the traffic. 
Donoho et al.~\cite{DoFlShPaCoStL:02} studied what happens if the stepping stone modifies its flow to evade detection with a maximum tolerable delay constraint. Then, with large enough sequences, they can correlate the traffic regardless of the modification. They use wavelets to separate the short-term behavior from the long-term behavior, and use the correlation on the latter. 
Blum et al.~\cite{BlSoVe:04} studied stepping-stone detection under a maximum tolerable delay constraint. They count the difference between the number of packets in both flows. When this difference goes over a certain value they conclude that the flows are not correlated.

The common drawback of those three methods is that they require a large number of packets to achieve an acceptable performance.  This number can be significantly reduced by using active watermarks, which are discussed next. 

\subsection{Active watermarks}
Wang and Reeves~\cite{WaRe:03}  proposed the first active flow watermark. The watermark is  embedded in the IPDs. They first quantize the IPD and embed one bit of information by adding half of the quantization step or not. They argue that with sufficient redundancy (infinitely large watermark) the watermark can always be detected even if a timing perturbation is added to each packet. Hence, the drawback of this method is the amount of packets needed to obtain a good performance.
Wang et al.~\cite{WaChJa:07} proposed an interval centroid based watermark (ICB). They divide the time into intervals. In each interval they embed one bit of the watermark; if the bit is 0 they send a request in the first half of the interval, and if the bit is 1 they do it in the second half. Each bit is decoded according to which half of the interval the centroid falls in.
Yu et al.~\cite{YuFuGrXuZh:07} proposed an interval watermark based on Direct Sequence Spread Spectrum (DSSS) communication techniques in order to hide it. The DSSS signal is embedded by modifying the traffic rate. This method again requires a long sequence.

Houmansadr et al.~\cite{HoKiBo2:09} proposed RAINBOW, a non-blind watermark which is robust to packet drops and repacketization. They record the IPD, then they embed the watermark by modifying the IPDs by a different quantity ($+a$ for 1, $-a$ for 0, or vice versa). The normalized correlation is used in detection, and a selective correlation when dealing with added and dropped packets.
Houmansadr and Borisov~\cite{HoKi:11} proposed SWIRL (Scalable Watermark that is Invisible and Resilient to packet Losses). The flow is divided into intervals: half of them are used to determine the slots pattern and the other half  are used to actually embed the watermark by delaying packets so as they fall into certain slots.
Pyun et al.~\cite{PyPaReWaNi:12} proposed an interval-based watermark (IB) designed to resist attacks that modify the number of packets, such as flow splitting, chaff packets and repacketization. The information is embedded in the difference between the number of packets in two contiguous intervals. This method has the drawback of being more detectable compared to the others. 

All the methods discussed above have been shown to be detectable, see next section. 
This would  allow an attacker to modify the known-watermarked flows in such a way that the watermark is removed.

\subsection{Detecting watermarks}

Peng et al.~\cite{PeNiRe:06} showed how a watermark can be detected and replicated. They detect which packets do not come from the assumed one-way packet delay distribution. Using that information, they can recover the parameters of the watermark algorithm thus being able to replicate it. Specifically, they applied their attack against the watermark in~\cite{WaRe:03}.

Kiyavash et al.~\cite{HoKiBo:09} discovered how one can detect not only the watermark but also extract the parameters and the key, with several network flows watermarked using the same key. This attack is effective against most of the interval based watermarks: ICB, DSS and IB. However, RAINBOW and SWIRL are designed to be immune to this attack.

Luo et al~\cite{LuZhZhPeLe:11} showed that any practical timing-based traffic watermark causes noticeable alterations in the intrinsic timing features typical of TCP
flows, and so it can be easily detected. Concretely, they propose metrics based on the round-trip time (RTT), IPDs, and one-packet bursts, that can expose IB,  ICB, RAINBOW and SWIRL watermarks for any kind of traffic: bulk or interactive. 
Lin and Hopper~\cite{LiHo:12} proposed more efficient ways to deal with passive detection than \cite{LuZhZhPeLe:11}.
They also argued that security against passive detection is not sufficient, as a stronger adversary that knows the previous flow is feasible in many scenarios.

\section{Proposed Scheme} \label{sec:sc}
This section introduces the notation we use and explains how we correlate the flows to decide whether they are linked or not.

\begin{figure}
  \centering
    \includegraphics[width=0.8\columnwidth]{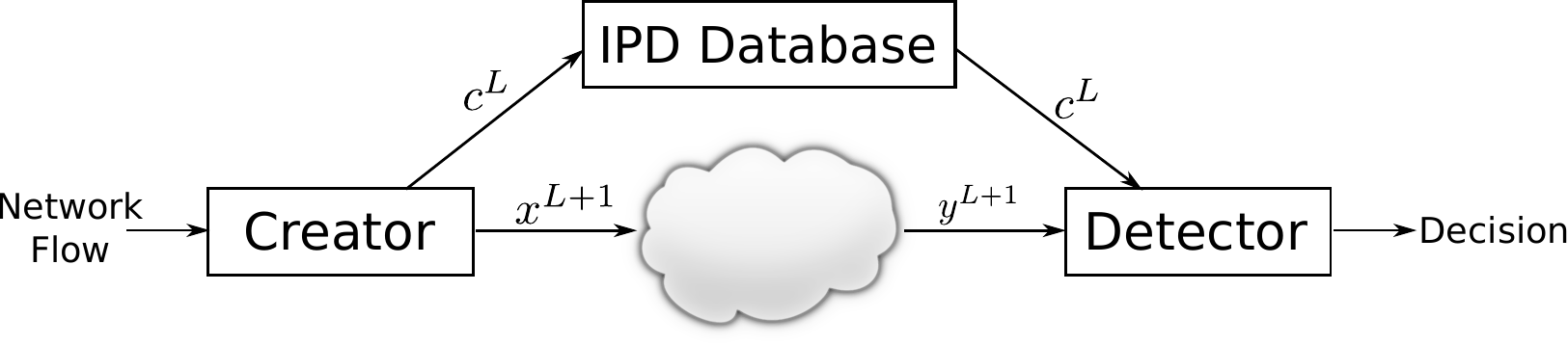}
  \caption{System Model.}
  \label{fig:sch}
\end{figure}

\textbf{Notation} We use the following notation. Random variables (r.v.) are denoted by capital letters (e.g., $X$), and their actual values by lower case letters (e.g., $x$). Sequences of $n$ random variables are denoted with a superscript (e.g., $X^n=(X_1,X_2,\dots, X_n)$). The  probability distribution function (pdf) of a continuous random variable $X$ is denoted by $f_X(x)$. When no confusion is possible, we drop the subscript in order to simplify the notation. The sample mean is denoted by $\bar{x}$. We summarize the name for each r.v. and parameters introduced in the sequel in Table \ref{tab:nam}.

\subsection{System model}
\begin{table}
\centering
\begin{tabular}{c l}
\multicolumn{2}{c}{\textbf{Random Variables}}\\
$A$ & PDV modification by an Attack\\
$C$ & IPD at the Creator\\
$D$ & IPD at the Detector\\
$J$ & PDV or Jitter introduced by the network\\ 
$N$ & Network delays\\
$X$ & Timing information at the creator\\
$Y$ & Timing information at the detector\\
$W$ & IPD modification by a Watermark\\ 
\multicolumn{2}{c}{\textbf{Parameters}}\\
$L$ & Original sequence length \\
$M$ & Matched sequence length \\
$\eta$ & Threshold for the likelihood-ratio test to reject $H_0$\\
$\rho$ & Synchronization constant\\
$\gamma$ & Threshold to consider a packet as lost\\
$P_L$ & Packet loss probability\\
$P_{NL}$ & Probability of packet loss due to the network\\
$P_M$ & Probability of not detecting a packet\\
$S$ & Number of subflows after a split\\
$A_{max}$ & Maximum delay constraint for an attacker\\
$W_{max}$ & Maximum delay constraint for the watermark\\
\end{tabular}
\caption{Random variables and constants used.}
\label{tab:nam}
\end{table}

Figure \ref{fig:sch} illustrates our system model. A flow of length $L+1$ packets, that we are interested in tracking, goes through a certain link, termed ``creator'', where we can measure its packet timing information, $X^{L+1}$. The $i$th inter-packet delay (IPD) at the creator is defined as $C_i=X_{i+1}-X_{i}, \;i=1,\dots L$, and these values are saved for later use in detection. This flow continues through the network without any modification.

 The ``detector'' is another link in which we can measure the timing information, $Y^{L+1}=X^{L+1}+N^{L+1}$, where $N_i$ is the network delay suffered by the $i$th packet. Then, the IPDs at the detector are
\begin{align}\label{eq:ipd}
D_i&=Y_{i+1}-Y_i=X_{i+1}-X_{i}+N_{i+1}-N_i \nonumber \\
&=C_i+J_i, \; \ i=1\dots L
\end{align}
where $J_i=N_{i+1}-N_i$ represents the packet delay variation (PDV), also known as jitter.

By using the information of the actual values $c^L$ and $d^L$, the detector has to decide correctly if the two flows are linked. Two flows are linked if they follow a common timing pattern due to sharing the same source (i.e. the unencrypted payload is the same). Formally, we can express this problem via classical hypothesis testing with the following hypotheses:
\begin{align*}\label{eq:hypothesis}
	\text{H}_0\text{:} & \; \text{ The flows are not linked.}\\
	\text{H}_1 \text{:}& \; \text{ The flows are linked.}
\end{align*}

\subsection{Performance Metrics}
To measure performance, we use two metrics: the probability of detection ($P_D$), which represents the probability
of deciding that the flows are linked when they actually are; and the probability of false positive ($P_F$),
which represents the probability of deciding incorrectly that the flows are linked. Formally, $P_D$ is the
probability of deciding $H_1$ when $H_1$ holds, whereas $P_F$ is the probability of deciding $H_1$ when $H_0$ holds.

Typically, performance is graphically represented using the so-called
ROC (Receiver Operating Characteristic) curves, which represent
$P_D$ vs. $P_F$. In a practical setting, one fixes a certain
value of $P_F$ (that has to be very small if we want to achieve
a high reliability) and then measure $P_D$ (which we would
like to be as large as possible).

In order to compare different ROCs in a simple way, we use the AUC (area under the ROC curve), 
a measure that takes a value of 1 in the case of perfect detection and 0.5 in case of random choice. 
The AUC is shown in the legend of each graph.

\section{Basic Detector}\label{sc:det1}
In this section we derive our detector and model the distributions of PDVs and IPDs as needed.

\subsection{Detector construction}
In order to obtain the best possible performance, we construct the optimal detector, which is the likelihood ratio
test. Neyman-Pearson lemma proves that this test is the most efficient one between two simple hypotheses~\cite{NePe:33}. Hence, our detector chooses $H_1$ when
\begin{equation} \label{eq:det1}
\Lambda(d^{L},c^{L})=\frac{\mathcal{L}(H_1|d^{L},c^{L})}{\mathcal{L}(H_0|d^{L},c^{L})}=\frac{f(d^L|c^L,H_1)}{f(d^L|c^L,H_0)}>\eta.
\end{equation}
and $H_0$ in the opposite case. $\mathcal{L}$ represents the likelihood function and $\eta$ is a threshold that we fix to achieve a certain probability of false positive. 

Recall from (\ref{eq:ipd}) that if $H_1$ holds, then $D^L=C^L+J^L$. Conversely, if $H_0$ holds, $D^L$ is a sequence with joint pdf $f_{D^L}(d^L)$.

For feasibility reasons, we constraint the detector to use first-order statistics, discarding the information carried by higher-order statistics. This is equivalent to assuming sample-wise independence in the sequences $J^L$ and $D^L$. In Section \ref{sec:imp} we quantify the impact of this assumption on performance, comparing the real results with those that would be obtained for independent and identically distributed (i.i.d.) sequences. Under these assumptions, the likelihood ratio becomes
\begin{equation}\label{eq:detpr}
\Lambda(d^{L},c^{L})=\prod_{i=1}^L\frac{f_J(d_i-c_i)}{f_D(d_i)}.
\end{equation}

Therefore, we need to model the PDVs and the IPDs, i.e. determine $f_J(j)$ and $f_D(d)$.

\subsection{Modeling the packet delay variation}\label{sec:PDV}
To model the distribution of the PDVs, we first measure them in several real connections, then fit these data to some candidate distributions and select the distribution that matches best.

The measured delays are reported  in \cite{ElPe:13}. This dataset contains the delays between two hosts during 72 hours, and for 11 different scenarios. As it is customary, we separate these data into three subsets: training, validation and test, using 24 hours of data for each.

Scenarios 1 to 9 measure common Internet connections between two hosts. 
Scenario 10 models the delays of a stepping-stone scenario, where a host in Oregon is retransmitting  to a host in California the flow coming from a host in Virginia.
Scenario 11 measures the delays associated with one instance of the Tor network~\cite{DiMaSy:04}.
In order to get a general idea about the connection scenarios, we show some basic information of the hosts and the connections in Table~\ref{tab:delst}, where $P_{NL}$ is the probability of packet loss and the source and destination are represented with ISO 3166 codes~\cite{ISO:1998}.

\begin{table}
\centering
\begin{tabular}{l c c c c}
 & Source & Dest. &  $\bar{n}$ [ms] & $P_{NL}$\\
Sc1 & CA-US & NM-US & $15.4$ & $1.9 \cdot 10^{-4}$ \\
Sc2 & OR-US & NM-US & $26.7$ & $6 \cdot 10^{-4}$ \\
Sc3 & VA-US & NM-US &$41.4$ & $1.7 \cdot 10^{-3}$ \\
Sc4 & ES & NM-US & $93.6$& $0$ \\
Sc5 & IE & NM-US & $73.6$ & $9.8\cdot 10^{-5}$\\
Sc6 & JP & NM-US & $69.4$ & $1.0 \cdot 10^{-4}$ \\
Sc7 & AU & NM-US & $109.4$ & $1.2 \cdot 10^{-3}$ \\
Sc8 & BR & NM-US & $88.9$ & $9.5 \cdot 10^{-4}$\\
Sc9 & SG & NM-US & $110.5$& $9.4 \cdot 10^{-2}$\\
Sc10 & VA-US & CA-US & $63.1$ & $5 \cdot 10^{-3}$\\
Sc11 & NM-US & NM-US & $3117.2$ & $0.16$\\
\end{tabular}
\caption{Basic Statistics of the measured delays.}
\label{tab:delst}
\end{table}

From these measured delays we calculate the measured PDV as $j_i = n_{i+1} - n_i$. The basic statistics from Table \ref{tab:PDVst} imply a nearly symmetric (i.e., small skewness) and leptokurtotic distribution (i.e., sharp peak and heavy tail).

\begin{table}
\centering
\begin{tabular}{l c c c c}
 & $\bar{j}$ [\text{s}] & Var. $[\text{s}^2$] & Skew. & Kurtosis \\
Sc1 & $1 \cdot 10^{-10}$ & $1 \cdot 10^{-4}$ & $0.02$ &$83185$\\
Sc2 & $-2 \cdot 10^{-10}$ & $1 \cdot 10^{-5}$ &$5.84$ & $408$\\
Sc3 & $-2 \cdot 10^{-10}$& $2 \cdot 10^{-3}$&$0.003$ &$85187$\\
Sc4 & $1 \cdot 10^{-9}$& $1 \cdot 10^{-6}$ & $17.1$ &$81139$\\
Sc5 & $-7 \cdot 10^{-9}$& $3 \cdot 10^{-6}$& $3.81$ & $622$\\
Sc6 & $-2 \cdot 10^{-9}$& $6 \cdot 10^{-5}$& $0.78$ & $71212$\\
Sc7 & $2 \cdot 10^{-8}$& $2 \cdot 10^{-5}$& $-0.01$ & $78821$\\
Sc8 & $9 \cdot 10^{-9}$& $6 \cdot 10^{-3}$& $ -10^{-5}$ &$19893$\\
Sc9 & $2 \cdot 10^{-8}$ & $4 \cdot 10^{-6}$&$4.41$& $620$\\
Sc10 & $-8 \cdot 10^{-9}$ & $3 \cdot 10^{-4}$&$2.37$ &$22789$ \\
Sc11 & $-1 \cdot 10^{-6}$ & $6 \cdot 10^{-3}$&$2.97$ &$410$ \\
\end{tabular}
\caption{Basic Statistics of the measured PDV.}
\label{tab:PDVst}
\end{table}

To construct the model, we make the same assumptions as to build the test, i.e. an i.i.d. sequence.
The candidate distributions were selected among the ones that have support on $\mathbb{R}$ and possess the mentioned characteristics. The chosen distributions
are Cauchy, Gumbel, Laplace, Logistic and Normal. Their pdfs are summarized in Table \ref{tab:pdf}, where the indicator function $\indicator{[a,b]}(x)$ takes the value 1 when $x \in [a,b]$, and is 0 otherwise.

\begin{table}
\centering
\begin{tabular}{ l  c  }
\textbf{Distrib.} & \textbf{pdf} \\
\hline
Cauchy& $f(x|{\mu, \sigma})=\frac{1}{\pi \sigma\left( 1+ (\frac{x-\mu}{\sigma})^2 \right)}$\\
\hline
Gumbel & $f(x|{\mu, \sigma})=\frac{1}{\sigma} exp\left( -\frac{x-\mu}{\sigma}-exp\left( -\frac{x-\mu}{\sigma}\right)\right)$\\
\hline
Laplace& $f(x|{\mu, \sigma})=\frac{1}{2\sigma} \exp{-\frac{|x-\mu|}{\sigma}}$\\
\hline
Logistic & $f(x|{\mu, \sigma})=\frac{ \exp\left( -\frac{x-\mu}{\sigma}\right)}{\sigma\left(1+exp\left( -\frac{x-\mu}{\sigma}\right)\right)^2} $\\
\hline
Normal& $f(x|{\mu, \sigma})=\frac{1}{\sigma\sqrt{2 \pi}} exp\left(-\frac{(x-\mu)^2}{2\sigma^2}\right)$\\
\hline
Exp. & $f(x|{\lambda})=\lambda \exp(-\lambda x) \indicator {[0, \infty)}(x)$\\
\hline
Pareto & $f(x|{\alpha, x_m})= \frac{\alpha x_m^\alpha}{x^{\alpha+1}} \indicator {[x_m, \infty)}(x)$\\
\hline
LogNor.& $f(x|{\mu, \sigma^2})=\frac{1}{x\sqrt{2 \pi \sigma^2}} exp\left(-\frac{(\log x-\mu)^2}{2\sigma^2}\right)$\\
\hline
LogLog. & $f(x|{\alpha, \beta})=\frac{(\beta/\alpha)(x/\alpha)^{\beta-1}}{\left(1+(x/\alpha)^\beta\right)^2}\indicator {[0, \infty)}(x)$\\
\hline
Weibull & $f(x|{\gamma, \beta})= \frac{\gamma}{\beta} x^{\gamma-1} \exp\left( -\frac{x^\gamma}{\beta}\right)\indicator {[0, \infty)}(x)$\\
\end{tabular}
\caption{PDFs of the candidate distributions for PDV and IPD.}
\label{tab:pdf}
\end{table}

We estimate the respective parameters using robust statistics, to prevent that outliers affect the measures. These estimators are based on the median and median absolute deviation and calculated as explained in~\cite[Chapter~3]{Ol:08}. Afterwards, we measure the goodness of fit between the validation sequence and the model using the square root of the Jensen-Shannon divergence (JSD), $D_{JS}$~\cite{EnSc:03}. This is a metric for two probability densities $P, Q$, which is based on the Kullback-Leibler divergence (KLD) as follows:
\begin{equation}
D_{JS}(P,Q)=\sqrt{\frac{1}{2}\left(D(P\mid \mid M)+D(Q\mid \mid M)\right)}
\end{equation}
where $M = \frac{1}{2} (P+Q)$ is the mid-point measure, and $D(\cdot || \cdot)$ is the KLD, defined as
\begin{equation}
D(P,Q)=\sum_{i \in \mathcal{P}} f_P(i) \log {\left(\frac{f_P(i)}{f_Q(i)}\right)}
\end{equation}

\begin{table}
\centering
\begin{tabular}{ l c c c c c}
Scenario & Cau. & Gum.& Lap.& Log.& Nor. \\
Sc. 1& $0.168$ & $0.218$ & $0.101$ & $0.123$ & $0.159$\\
Sc. 2& $0.156$ & $0.201$ & $0.157$ & $0.150$ & $0.171$\\
Sc. 3& $0.135$ & $0.211$ & $0.163$ & $0.167$ & $0.192$\\
Sc. 4& $0.294$ & $0.369$ & $0.252$ & $0.270$ & $0.296$\\
Sc. 5& $0.153$ & $0.193$ & $0.139$ & $0.135$ & $0.159$\\
Sc. 6& $0.203$ & $0.174$ & $0.152$ & $0.120$ & $0.130$\\
Sc. 7& $0.136$ & $0.300$ & $0.231$ & $0.267$ & $0.298$\\
Sc. 8& $0.168$ & $0.307$ & $0.195$ & $0.261$ & $0.308$\\
Sc. 9& $0.183$ & $0.185$ & $0.171$ & $0.141$ & $0.146$\\
Sc. 10& $0.227$ & $0.384$  & $0.340$ & $0.364$ & $0.384$\\
Sc. 11& $0.251$ & $0.201$  & $0.194$ & $0.228$ & $0.253$\\
\end{tabular}
\caption{Goodness of fit of the candidate distributions for PDV.}
\label{tab:godPDV}
\end{table}

Results from Table \ref{tab:godPDV} show that no distribution stands out above the rest,  being the Laplace and the Cauchy distributions the best fits. 

The Laplacian is the most commonly used model for the jitter, but Rio-Dominguez et al.~\cite{RiMuToVa:10} claimed that an alpha-stable distribution models  it better. Note that a Cauchy distribution is a particular case of an alpha-stable distribution, but we do not generalize it further, as we are interested in a close-form pdf model.

The performance of the two possible detectors, based on Laplace and Cauchy distributions, respectively, is evaluated in Section \ref{sec:imp}.
 
\subsection{Modeling the Inter-Packet Delays}
In many works it is assumed a Poisson model for the traffic because of its desirable theoretical properties~\cite{Kl:75}. This model implies that IPD times are an i.i.d. exponentially distributed sequence. But Paxson et al.~\cite{PaFl:95} have shown that this model is not accurate in interactive applications.

We model the IPDs on both SSH and HTTP protocols. As done in~\cite{PaFl:95}, we only take into account packets that are separated at least by 10 ms, considering that if two packets are separated by less than 10 ms they are subpackets of the same packet. Therefore, the considered IPDs are lower bounded by 10 ms. We use the captures from Dartmouth College~\cite{KoHeAbYe:04}, using the traces from Fall 03 as training set, Spring 02 as validation set and Fall 01 as test set for the simulator. The basic characteristics of these sets are shown in Table \ref{tab:sets}.

\begin{table}
\centering
\begin{tabular}{l c c}
Set &Flows &Packets\\
SSH Train. & $6447$ & $14442323$\\
SSH Val. & $1128$ & $2594550$\\
SSH Sim. & $714$ & $16595655$\\

HTTP Train. & $1108909$ & $356620487$\\
HTTP Val. & $208896$ & $63982082$ \\
HTTP Sim. & $1007545$ & $322853437$\\
\end{tabular}
\caption{Characteristics of the IPD sets.}
\label{tab:sets}
\end{table}

We estimate the parameters through maximum likelihood estimation (MLE) and measure the goodness of fit using the square root of the JSD. The candidate distributions are: Exponential, Pareto, Log-Normal, Log-Logistic, and Weibull. Their pdfs can be seen in Table \ref{tab:pdf}.

\begin{table*}
\centering
\begin{tabular}{l c c c c}
Distribution & Error SSH &Par SSH &Error HTTP &Par HTTP\\
Exponential & $0.756$ & $\lambda=5.46$& $0.758$ & $\lambda=12.69$\\
Pareto & $0.149$ & $\alpha=0.86,\; x_m=10^{-2}$& $0.247$ & $\alpha=0.53,\; x_m=10^{-2}$\\
Log-Normal & $0.627$ &$\mu=-1.14,\; \sigma^2=1.43$ &$0.723$ &$\mu=-0.40,\; \sigma^2=4.02$\\
Log-Logistic & $0.343$ &$\alpha=0.27,\; \beta=1.77$ & $0.508$ &$\alpha=0.47,\; \beta=0.95$\\
Weibull &$0.554$ &$\gamma=0.49,\; \beta=0.81$ &$0.591$ &$\gamma=0.40,\; \beta=1.33$\\
\end{tabular}
\caption{MLE Estimator and goodness of fit of the candidate distributions for IPD.}
\label{tab:mlegodipd}
\end{table*}

Results shown in Table \ref{tab:mlegodipd} confirm the findings of Paxson et al., i.e., that the Pareto distribution is a better model for interactive traffic. In non-interactive traffic such as HTTP, this model also gives acceptable results. Therefore, we will assume that
\begin{equation}
f_D(d)=\alpha x_m^\alpha d^{-\alpha-1}\indicator{[x_m,\infty)}(d).
\end{equation}

\subsection{Detector}
Once we have a model for the IPD and PDV sequences, we derive the likelihood test.

If Cauchy distributed PDVs are assumed, the test chooses $H_1$ when 
\begin{align}\label{eq:test1C}
\Lambda(d^L,c^L)= \prod_{i=1}^L \frac{(d_i)^{\alpha+1}}{\pi \sigma \alpha x_m^\alpha \left(1+\left(\frac{d_i-c_i}{\sigma}\right)^2\right)}>\eta
\end{align}
and $H_0$ otherwise.

In the case that a Laplace model for PDV is adopted, then
\begin{align}\label{eq:test1L}
\Lambda(d^L,c^L)= \exp \left(-\frac{\sum_{i=1}^L|d_i-c_i|}{\sigma}\right) \frac{\left(\prod_{i=1}^L d_i\right)^{\alpha+1}}{(2\sigma \alpha x_m^\alpha)^{L}}.
\end{align}

\section{Performance}\label{sc:prf}
In this section we construct a simulator and present the scenarios we use in the remaining of the paper. Afterwards, we test the model assumptions and measure the performance with different sequence lengths.

\subsection{Simulator and Scenarios}
\begin{figure}
  \centering
    \includegraphics[width=0.8\columnwidth]{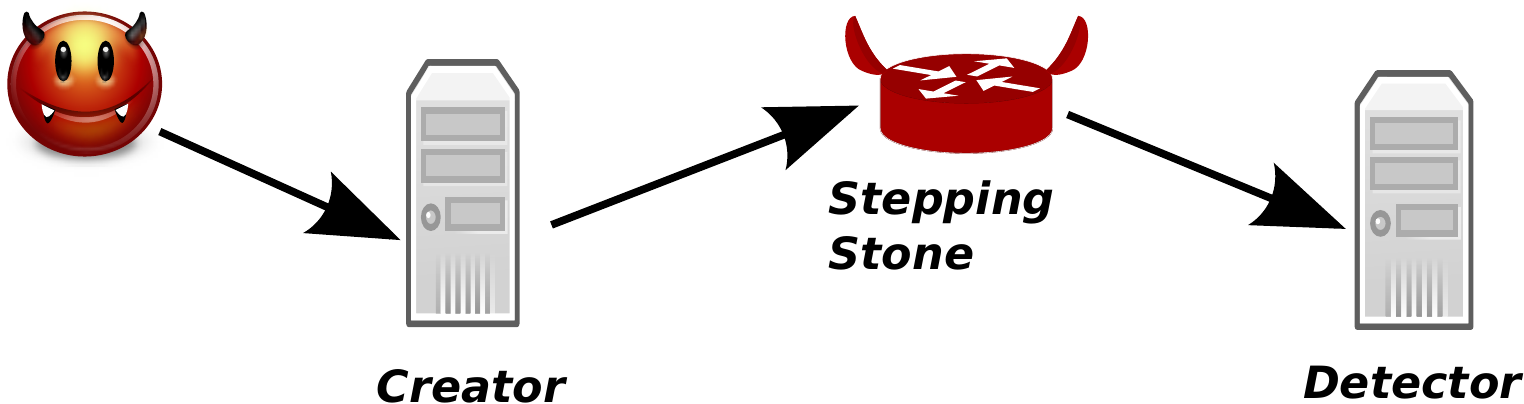}
  \caption{Simulated Scenario A.}
  \label{fig:simscA}
\end{figure}
\begin{figure}
  \centering
    \includegraphics[width=0.8\columnwidth]{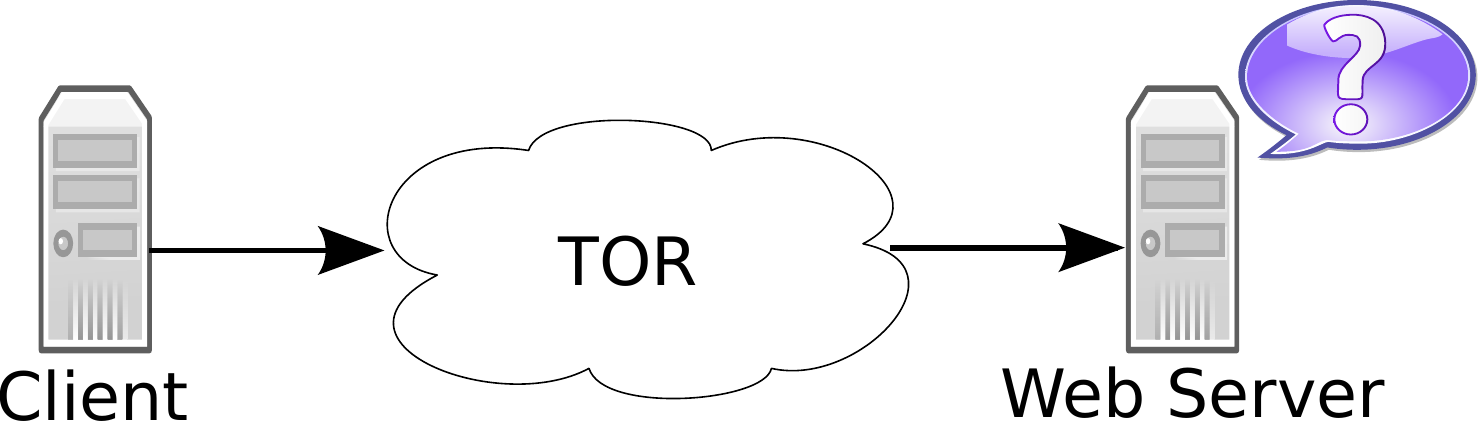}
  \caption{Simulated Scenario B.}
  \label{fig:simscB}
\end{figure}
Simulations are carried out in the following way. First, we generate timing information at the creator using the IPD test set, $X_1^{L+1}$. The purpose of this sequence is to evaluate the performance when $H_1$ holds. A delay is added to each packet using the measured delays from the test set (as explained in the following paragraphs), obtaining $Y_1^{L+1}$. 
We generate a second sequence $Y_0^{L+1}$, using the IPD test set; this sequence has the purpose of evaluating the performance under $H_0$. Finally, we use the Test from \eqref{eq:test1C} or \eqref{eq:test1L}, to obtain both $\Lambda(d_0^L, c_1^L)$ and $\Lambda(d_1^L, c_1^L)$. This experiment is repeated $10^6$ times, and for different values of $\eta$ we obtain $P_D$ as the rate of $\Lambda(d_1^L, c_1^L)>\eta$, and $P_F$  as the rate of $\Lambda(d_0^L, c_1^L)>\eta$. Note that due to the number of runs, $P_F<10^{-5}$ cannot be measured and results of this order are not accurate. 

The sequences are generated in the following way: we
place all the IPDs from the test set in an order-preserving list. The starting point is randomly selected from the list and the generated IPDs are the following $L$ values.

For generating the delays, we used the test set as a list with the delay every 50 ms. We select one value randomly from the list that will be considered time 0 ms; the following values will represent the delay at times 50 ms, 100 ms, and so on. To obtain the delays at times where we do not have a measure, we use linear interpolation.

The performance is evaluated in the two scenarios depicted in Figures \ref{fig:simscA} and \ref{fig:simscB}. Scenario A represents a stepping stone that forwards SSH traffic inside the Amazon Web Services~\cite{AWS} network. The creator, stepping stone and detector are EC2 instances located in Virginia, Oregon and California, respectively. This example corresponds to tracing the source of an attack that was launched from a compromised Amazon instance. The simulated delays correspond to those of Scenario 10 in Section \ref{sec:PDV}, where the standard deviation of the network delay is 4 ms.

Scenario B simulates a web page accessed from Tor network whose real origin is to be found, and where the creator will be the web page and the detector the client. For instance, this case can correspond to a company in whose forum an anonymous insulting post has been placed using Tor and it is to be known whether the source comes from an employee within the company. The simulated delays correspond to the measurements of Scenario 11 in Table \ref{tab:godPDV}, where the standard deviation of the network delay is 340 ms.

\subsection{Impact of our assumptions}\label{sec:imp}
\begin{figure}
  \centering
    \includegraphics[width=0.9\columnwidth]{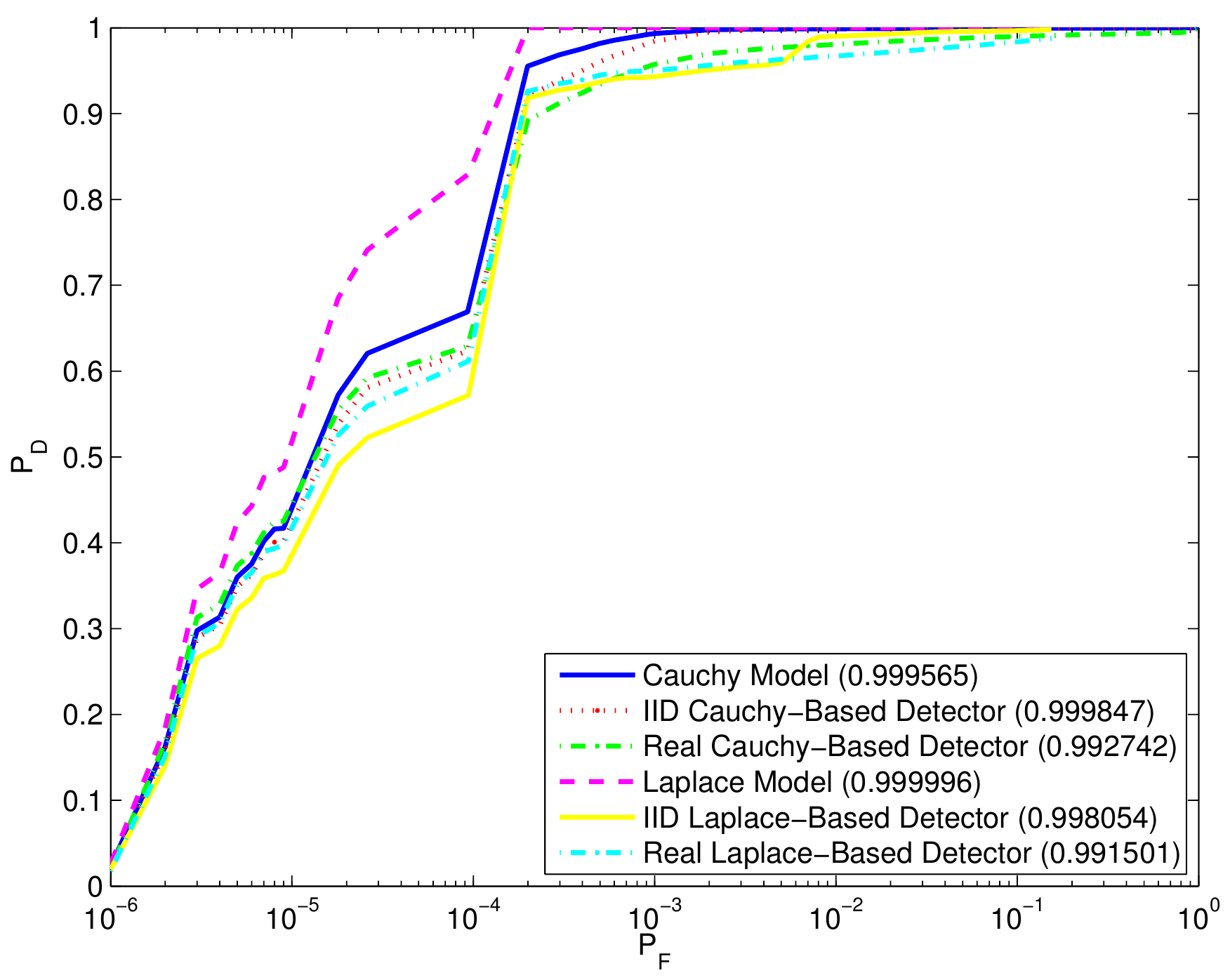}
    \caption{Impact of assumptions in Scenario A with $L=3$.}
  \label{fig:Sc1Imp}
\end{figure}
\begin{figure}
  \centering
    \includegraphics[width=0.9\columnwidth]{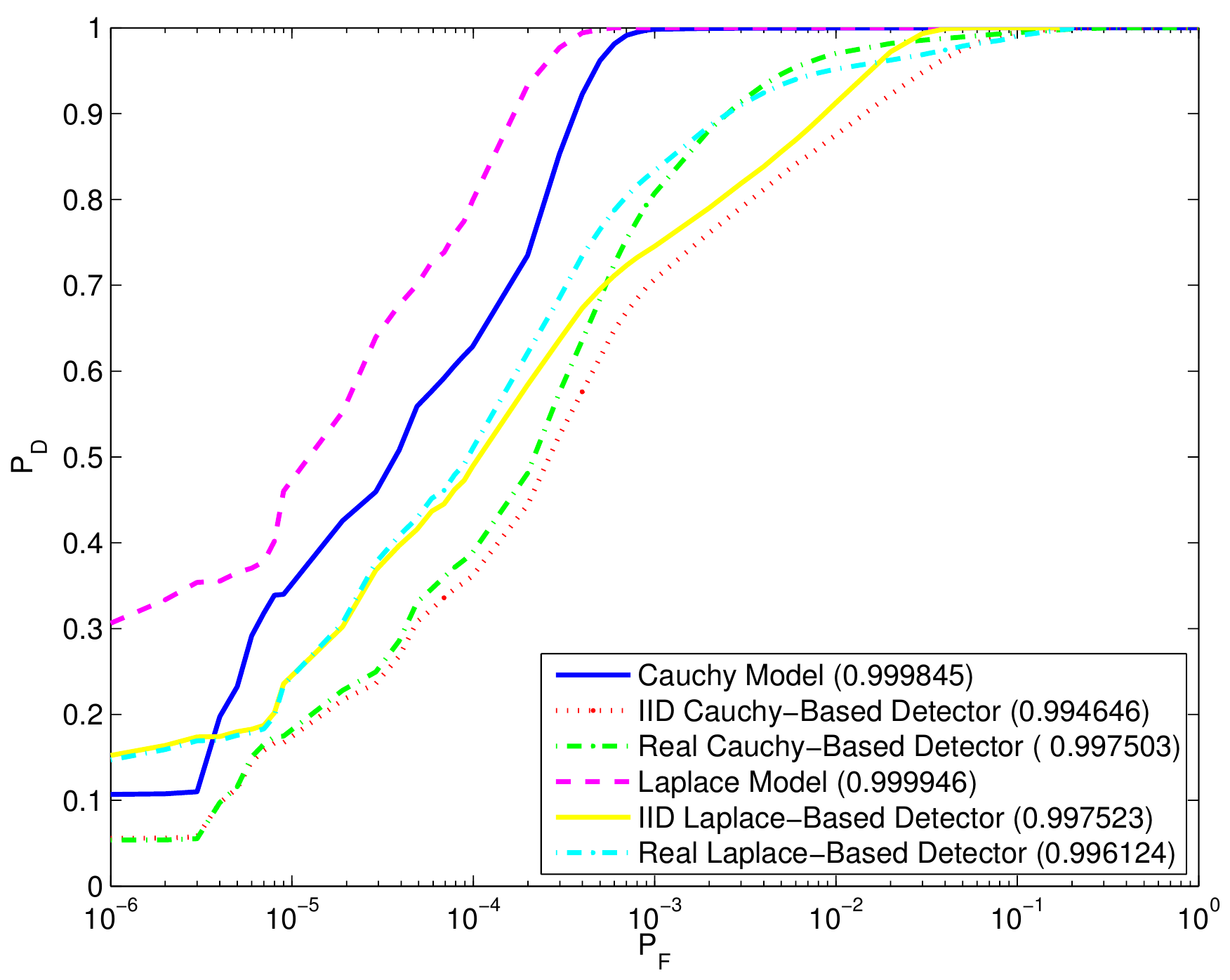}
    \caption{Impact of assumptions in Scenario B with $L=20$.}
  \label{fig:Sc2Imp}
\end{figure}

In this section, we wish to quantify the impact of the assumptions we have made, that is, the PDVs form an i.i.d. Cauchy or Laplace sequence. To this end, we extend our simulator to create 3 types of delays: first, according to the model (Cauchy or Laplace), second as a random sample from the data, and last, from the data maintaining the time correlation. $L=3$  is used for Scenario A and $L=20$ for Scenario B. Results are shown in Figures \ref{fig:Sc1Imp} and \ref{fig:Sc2Imp}. We notice two details: first, that the Cauchy-based detector gives slightly better performance than the Laplace under real data, and second, that the independence of the PDVs previously assumed slightly reduces the performance. In the sequel, we just derive the expressions for a Cauchy-based detector. The modification for a Laplace detector is rather straightforward. 

\subsection{Performance dependence on $L$}
\begin{figure}
  \centering
    \includegraphics[width=0.9\columnwidth]{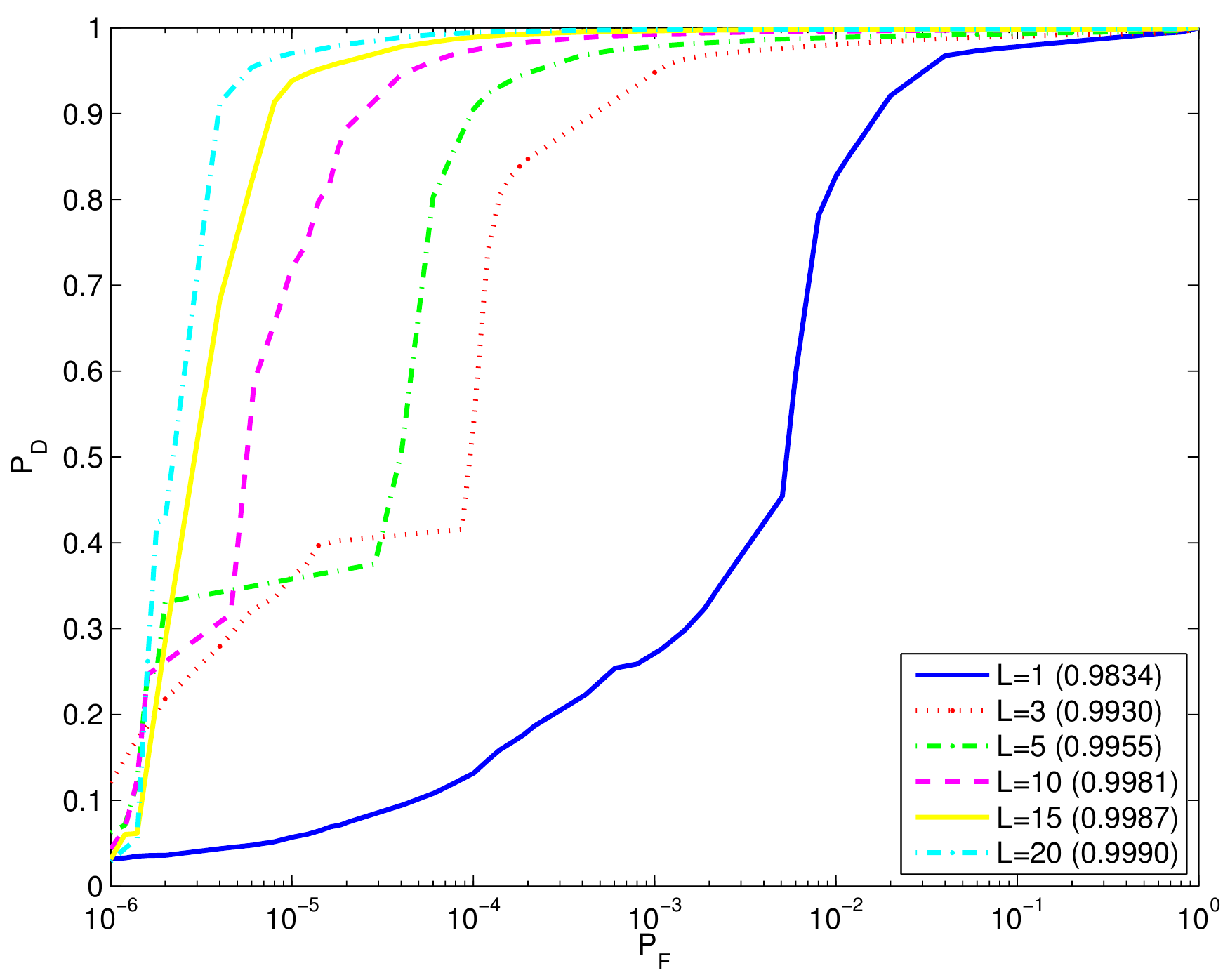}
  \caption{Performance  dependence with $L$ in Scenario A.}
  \label{fig:sim1L}
\end{figure}

\begin{figure}
  \centering
    \includegraphics[width=0.9\columnwidth]{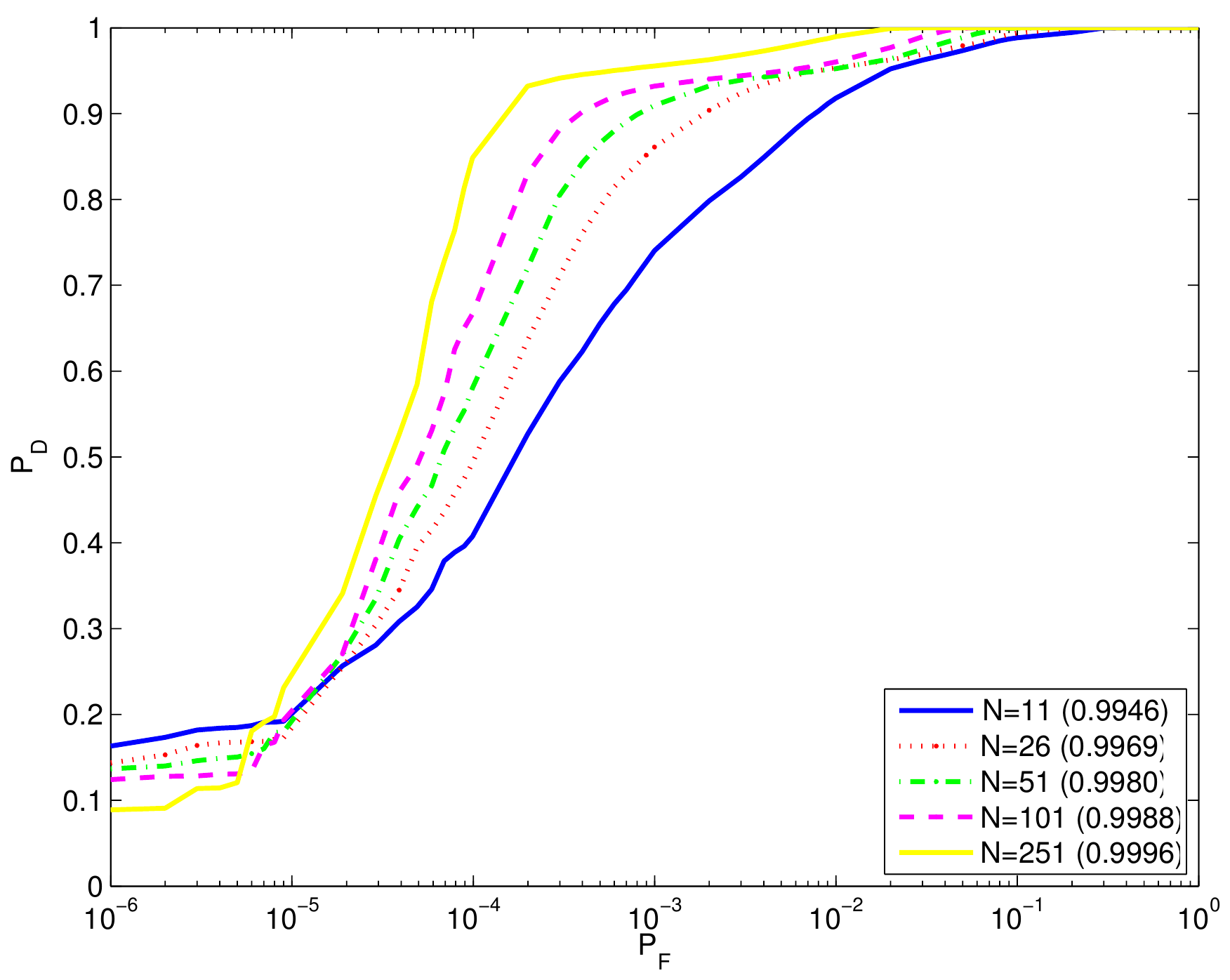}
  \caption{Performance dependence with $L$ in Scenario B.}
  \label{fig:sim2L}
\end{figure}

We want to evaluate how much performance is improved when longer sequences are used. The result is depicted in Figures \ref{fig:sim1L} and \ref{fig:sim2L}. We can see that Scenario B, whose IPDs have a larger variance because of the Tor network, needs much longer sequences to achieve the same performance. For instance, with fixed $P_F=10^{-4}$, in Scenario A for $L = 5$ we obtain $P_D=0.8926$. However, in Scenario B the $L$ needed for a comparable result is around $250$, with which we obtain $P_D=0.8947$. If we compare AUCs, in Scenario A with $L=5$ we obtain $0.9955$ while a similar result in Scenario B requires a value of $L$ between $10$ and $25$.

\section{Robust detector}\label{sc:det2}
The previous test does not take the existence of any countermeasure into account. Attacks to timing correlation can be exerted by introducing uncorrelated random delays, adding chaff traffic or splitting the flow, making the Test in \eqref{eq:test1C} ineffective. In this section, we build a test that is robust to these attacks. First, we deal with adding or removing packets from the flow, and then with random delays.

\subsection{Matching packets} 
Hitherto, we have assumed that there is a one-to-one relation between the flows at the creator and the detector; i.e., no packets are added or removed. This assumption is not necessarily valid for every situation, not only due to the presence of an active attacker, but also as a result of many applications that repacketize flows, changing the number of packets, for instance, SSH tunneling~\cite{rfc4251}.

To deal with packet addition and removal, we first choose the most likely packet at the detector for each packet at the creator. In the case that there is no packet likely enough, we consider the creator packet as lost.

Given the $i$th packet at the creator, we match it with the most
likely $j$th packet at the detector, denoting this as $i \rightarrow j$. Consequently, if $\rho$ is a synchronization constant to be discussed in Section~\ref{sec:syn}, and $\gamma$ is the threshold for which a packet is considered lost, the condition for a match in the $i$th packet is
\begin{equation}
|x_i-(y_j-\rho)| < |x_i-(y_k-\rho)|,\;\; \forall k\neq j,
\end{equation}
and to avoid considering it lost,
\begin{equation}
   |x_i-(y_j-\rho)|<\gamma.
\end{equation}

Threshold $\gamma$ should be large enough so that the probability $P_M$ that a packet is wrongly considered lost is very small, for instance, $10^{-6}$. Although this can lead to incorrectly matching with another packet when the packet is indeed lost,
the impact on Test \eqref{eq:test2} of this mismatch is very small. Empirically, the best performance we obtained for Scenario A is when $\gamma \approx 75$ ms and when $\gamma \approx 7$ s for Scenario B.

In practice, the standard deviation of the network delay can be larger than some of the IPDs, especially in Scenario 2, in which case the matching is likely to fail. The impact of these matching errors is evaluated in Section \ref{sec:adper}. In the case that most of the IPDs are smaller than the standard deviation of the network delay, a better matching function is the one used in~\cite{ElPe:13-3}. This corresponds to the injective function that minimizes the mean square error between $x^n$ and $y^m-\rho$, which has the drawback of a higher computational cost.

The matching process modifies the timing sequences to $x^{M+1}$ and $y^{M+1}$, where $M \leq L$, as the lost packets are removed. Formally, we can define the new sequences as $x^{M+1}=\{x_i\;|\;\exists\;j:\; i\rightarrow j\}$, and $y^{M+1}=\{y_j\;|\;\exists\;i:\; i\rightarrow j\}$.

\subsection{Test robust to chaff and flow splitting}

From \eqref{eq:test1C}, we can obtain a test robust to packet removal and insertion as
\begin{align}\label{eq:test2}
	\Lambda(d^M,c^M) =&  {P_L}^{L-M} \cdot \prod_{i=1}^{M} \bigg( P_L+(1-P_L) \cdot \nonumber\\
	&\left.\frac{(d_i)^{\alpha+1}}{\pi \sigma \alpha x_m^\alpha \left(1+\left(\frac{d_i-c_i}{\sigma}\right)^2\right)}\right),
\end{align}
where $P_L$ is the probability that a packet at the creator cannot be matched at the detector. This can be due to three reasons: network loss with a probability $P_{NL}$, lack of matching when the packet appears, and flow splitting  into $S$ subflows by the stepping stone, i.e., $(S-1)/S$ of the original packets  are not seen by the detector, as only one of the subflows traverses this link. Therefore, 
\begin{align}
P_L&=\frac{S-1+P_{NL}+P_M-P_{NL}P_M}{S} \approx \frac{S-1+P_{NL}+P_M}{S}
\end{align}

\subsection{Self-Synchronization}\label{sec:syn}
\begin{figure*}
        \centering
        \subfloat[Linked flows]{\includegraphics[width=0.45 \textwidth]{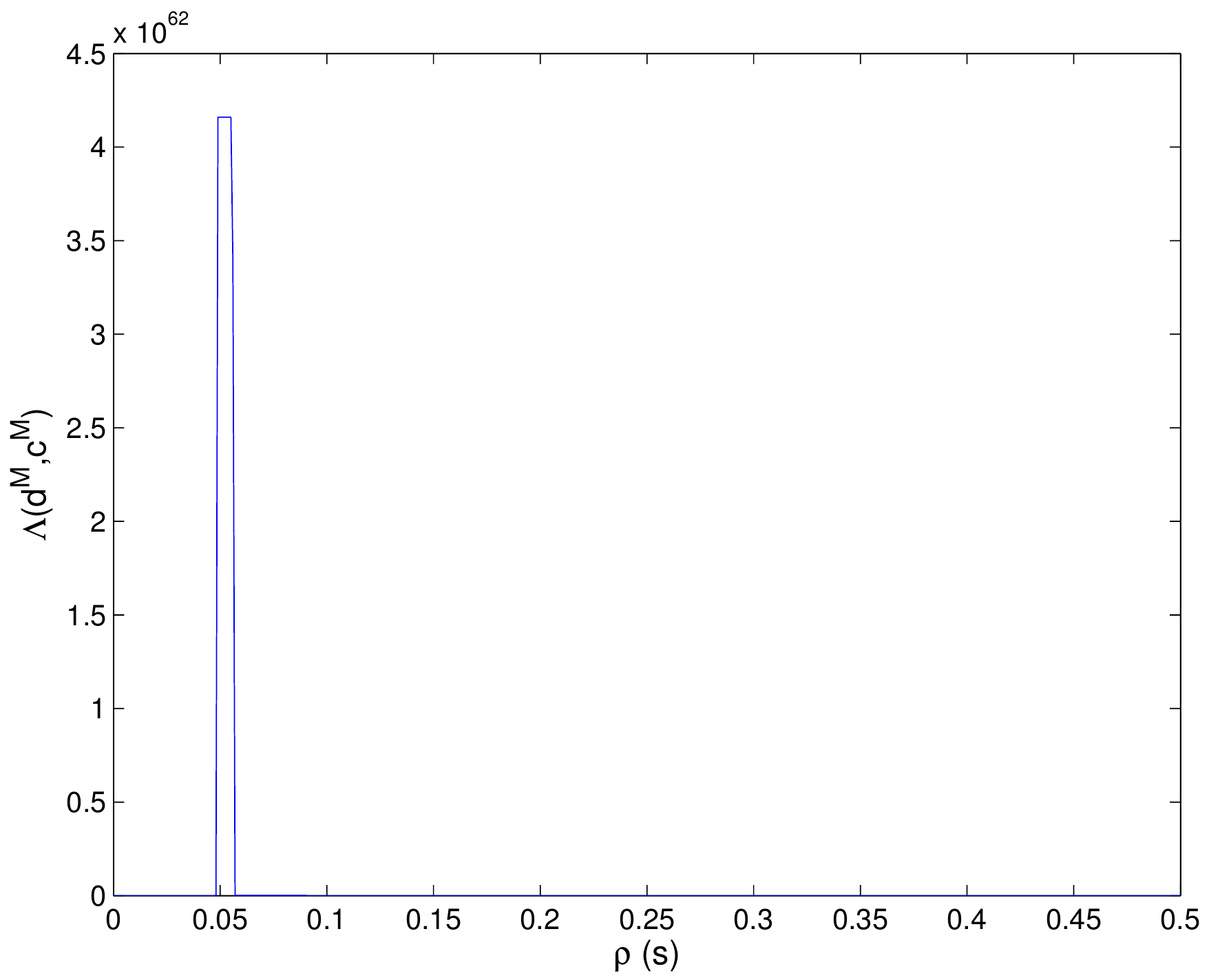}}\qquad
        \subfloat[Non linked flows]{\includegraphics[width=0.45\textwidth]{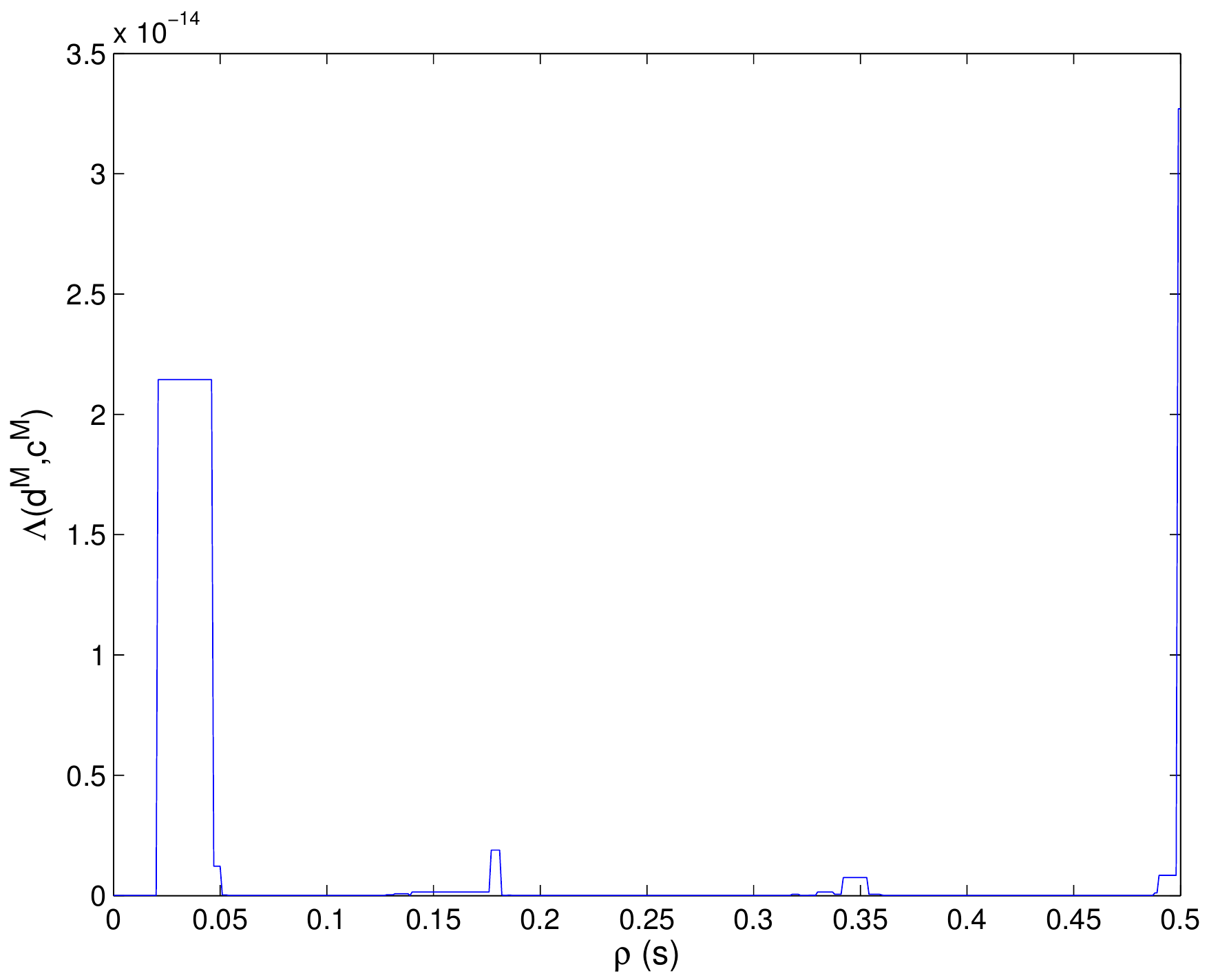}}
        \caption{Synchronization in Scenario A with $L=5$.}
        \label{fig:syn1}
\end{figure*}

We have mentioned that $\rho$ is a synchronization constant. The detector can perform detection maximizing the value of $\Lambda(d^M,c^M)$ with respect to $\rho$ through an exhaustive search. For instance, Figure \ref{fig:syn1} shows a detector trying values of $\rho$ using steps of 1 ms in the interval $[0,0.5]$ s. We can see that the maximum $\Lambda(d^M,c^M)$ occurs when $\rho \approx \bar{n}$, as expected. Recall that $\bar{n}$ is the sample mean of the network delays. 

\subsection{Robust test against random delays}
So far the situation where an attacker can inject random delays has not been considered. Random delay injection is a well-known technique for covert channel prevention and can be easily implemented via buffering by attackers across their step stones.

We assume that the attacker has the constraint of not being able to delay any packet more than $A_{max}$ seconds. Hence, she can modify the PDV by a quantity $A$ that falls in the interval $[-A_{max}, A_{max}]$. As we do not know the distribution of the attacker's random delay, the detector assumes a uniform distribution. Thus, the PDV at the decoder is $J'^M=J^M+A^M$, and
\begin{align}
 f_{J'}(j)=&\frac{1}{2 A_{max}\pi}\left((\arctan\left(\frac{j+A_{max}}{\sigma}\right)\right.\nonumber\\
 &\left.-\arctan\left(\frac{j-A_{max}}{\sigma}\right)\right)
\end{align} 
Consequently, the likelihood ratio becomes
\begin{align}\label{eq:test3}
	\Lambda(d^M,c^M) = & {P_L}^{L-M} \cdot \prod_{i=1}^{M} \bigg( P_L+(1-P_L) \cdot \nonumber\\
	&\left.\frac{(d_i)^{\alpha+1} f_{J'}(d_i-c_i)}{\alpha x_m^\alpha}\right)
\end{align}

A game-theoretic approach to this problem is taken in~\cite{ElPe:13-2}, where for simplicity the detector is constrained to estimating and compensating the attack. The optimal detector for the same game is derived in~\cite{ElPe:13-3}, where it is shown that a nearly deterministic attack impairs the detector more than a uniform distribution even if the detector knows the attack distribution.

\subsection{Performance}\label{sec:adper}

To evaluate the proposed robust algorithms, the functionalities of adding chaff traffic, splitting the flow, and delaying the packets randomly are implemented in our simulator. This is done as follows: each packet is delayed by a certain quantity. We implement two different delay strategies: a) the value is picked from a uniform distribution in the range $[0,A_{max}]$, and b) the values are taken to minimize \eqref{eq:test2}, i.e. the values are chosen by an intelligent adversary who knows both the test and its parameters. Then, the simulator adds traffic according to a Poisson process with a fixed rate proportional to the rate of the original traffic. Afterwards, it simulates the flow split, which is implemented by discarding packets as a Bernoulli process with a probability equal to $1-\frac{1}{S}$. Recall that $S$ is the number of subflows we divide the flow into.

\begin{figure}
  \centering
    \includegraphics[width=0.9\columnwidth]{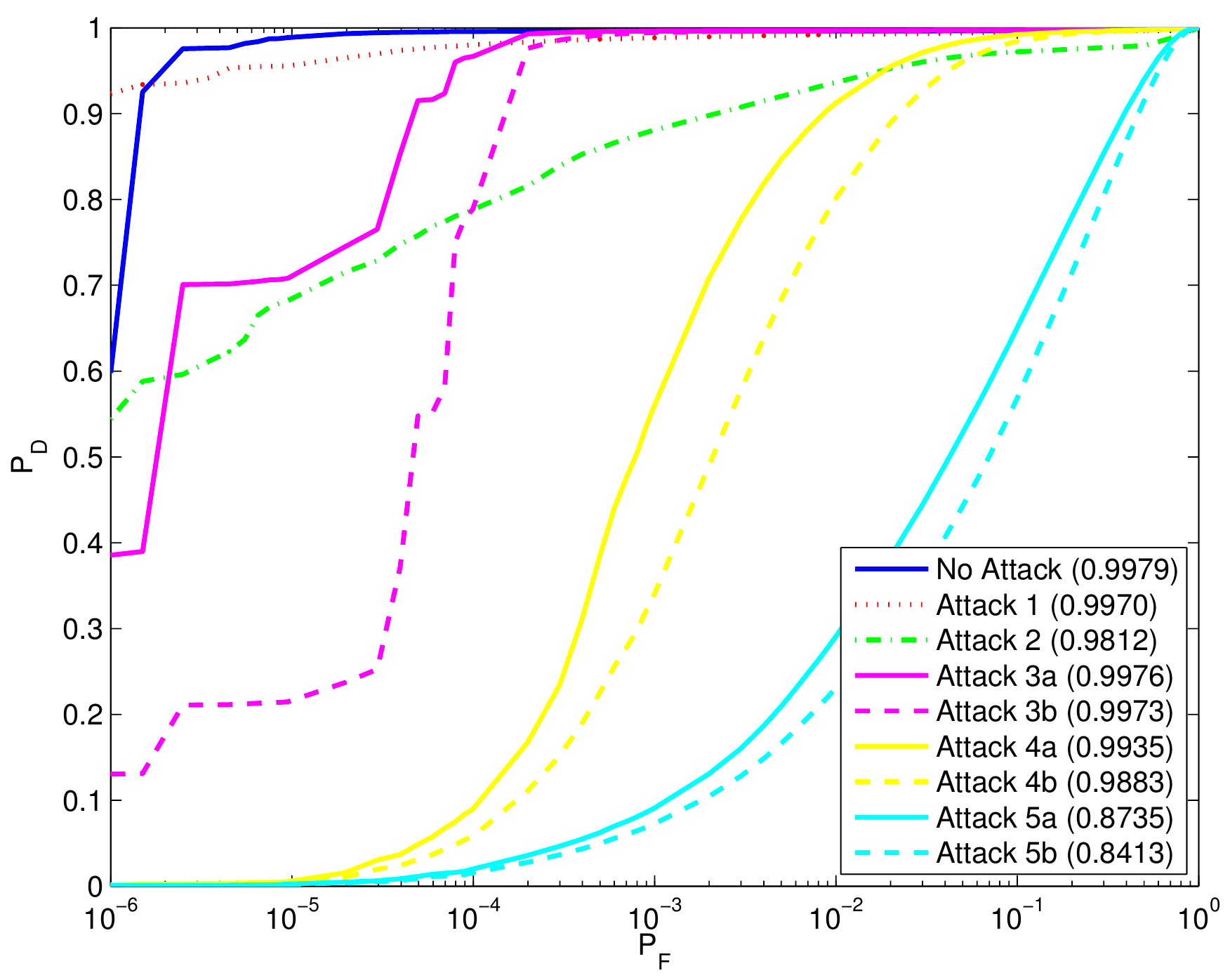}
  \caption{Performance under different traffic modifications in Scenario A, L=20.}
  \label{fig:att1}
\end{figure}

\begin{figure}
  \centering
    \includegraphics[width=0.9\columnwidth]{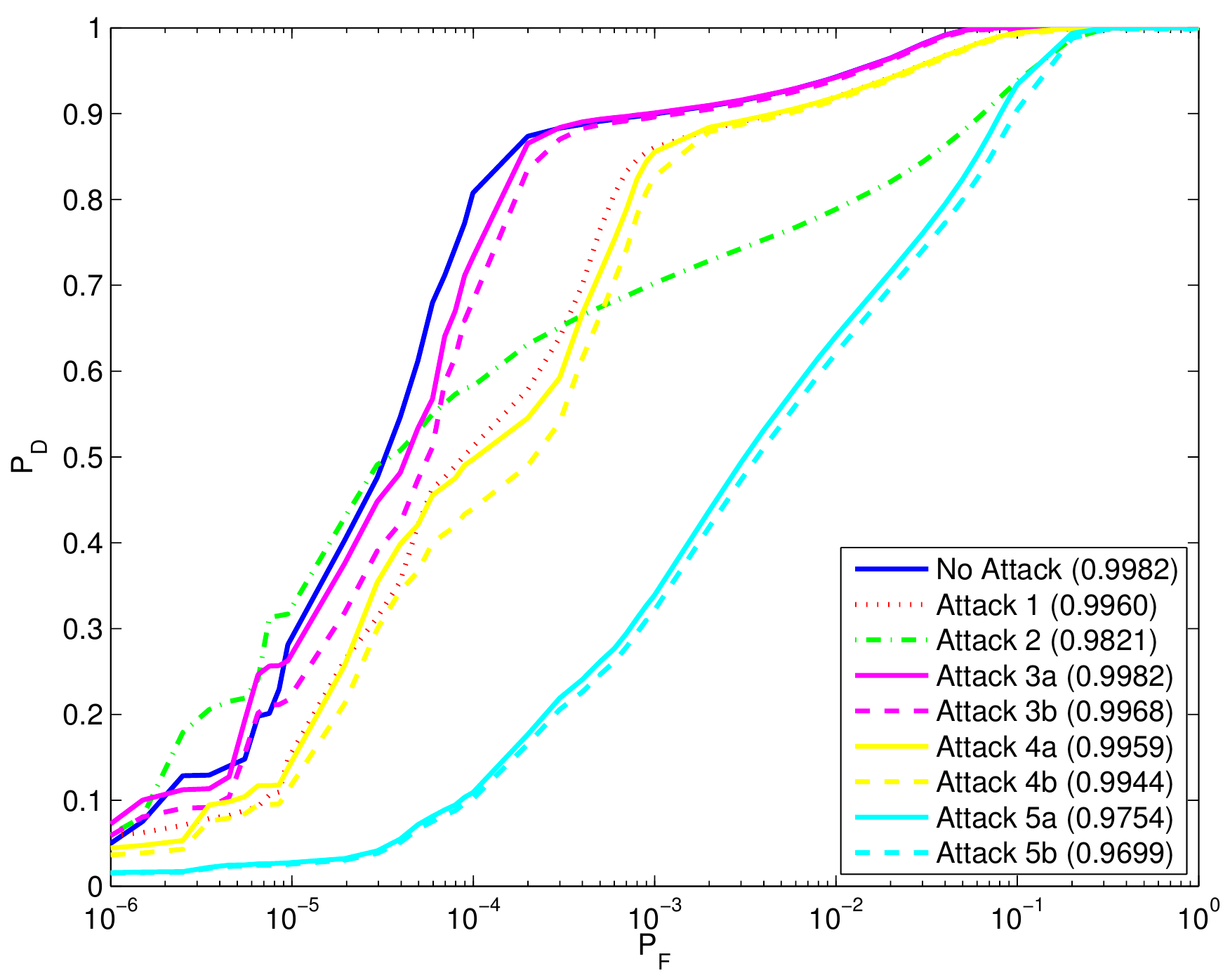}
  \caption{Performance under different traffic modifications in Scenario B, L=250.}
  \label{fig:att2}
\end{figure}

We created five different attacks. In the first three, we evaluate each traffic modification strategy separately, namely, Attack 1 adds 500\% of chaff traffic; Attack 2 splits the flow into 4 subflows; Attack 3 adds delays with $A_{max}=50$ ms; Attack 4 combines 500\% of chaff traffic with delays constrained to $A_{max}=50$ ms, and Attack 5 is a complex attack where a combination of Attack 4 with splitting the flow into 2 subflows takes place. For Attacks 3 to 5, we consider the two delay strategies specified above: with $Z$ indicating the attack number, we denote by $Za$ the case where the delays are chosen randomly, and by $Zb$ where they are chosen by an intelligent attacker. We simulate these situations using sequences of length $L=20$ in Scenario A and $L=250$ in Scenario B.
Results are depicted in Figures \ref{fig:att1} and \ref{fig:att2}.

Comparing these figures under no attacks with the corresponding plots for the case of no mismatches of Figs. \ref{fig:sim1L} and \ref{fig:sim2L}, we can evaluate the impact of mismatched packets, as the AUC drops from $0.9990$ to $0.9979$ in Scenario 1 and from $0.9996$ to $0.9982$ in Scenario 2.

In low jitter situations, namely Scenario A, chaff traffic by itself has little impact, but the effect when combined with random delays is significantly increased. The reason behind this is that in the first case the matching process chooses the real packets with a very low probability of error but when a random delay is added the probability of a mismatch increases. We also see that the flow splitting attack has a considerable impact as the received sequence length is reduced.

In high jitter situations, i.e. Scenario B, random delays have considerably smaller influence, because the standard deviation of the network delay is larger than the attack delay. In fact, due to the high network-delay variability, chaff traffic alone has a significant impact on performance without the need of an attacker injecting random delays.

\section{Comparison with an active watermark}\label{sc:watvspa}
We want to analyze how much performance can be improved by sacrificing undetectability. For this purpose, we create an active watermark designed with invisibility as a goal, and we study the trade off between performance and detectability.

We measure the latter as the KLD between the covertext, i.e., the sequence without watermark, and the stegotext, i.e., watermarked. Cachin~\cite{Ca:98} defines a stegosystem to be  $\epsilon$-secure against passive adversaries if $D(f_C||f_S)<\epsilon$, where $f_C$ is the distribution of the covertext and $f_S$ is the distribution of the stegotext. Hence, we measure the detectability as the minimum $\epsilon$ for which our system is $\epsilon$-secure.

The watermark is embedded adding a random uniform delay between $[0,W_{max}]$. Thus, the watermarked flow is $C'^L=C^L + W^L$, where $W^L$ is the embedded watermark which is triangular distributed between $[-W_{max},W_{max}]$  as it is the difference of two delays uniformly distributed. At the detector, we receive $D^L=C'^L+J^L$. The detector remains \eqref{eq:test3} using $c'^L$ instead of $c^L$.

We assume that the attacker knows the original traffic as done in \cite{LiHo:12, PeNiRe:06} and wants to test for the existence of a watermark. Therefore, the attacker's goal is to differentiate between $W^L+J^L$ and $J^L$. 

\begin{figure}
  \centering
    \includegraphics[width=0.9\columnwidth]{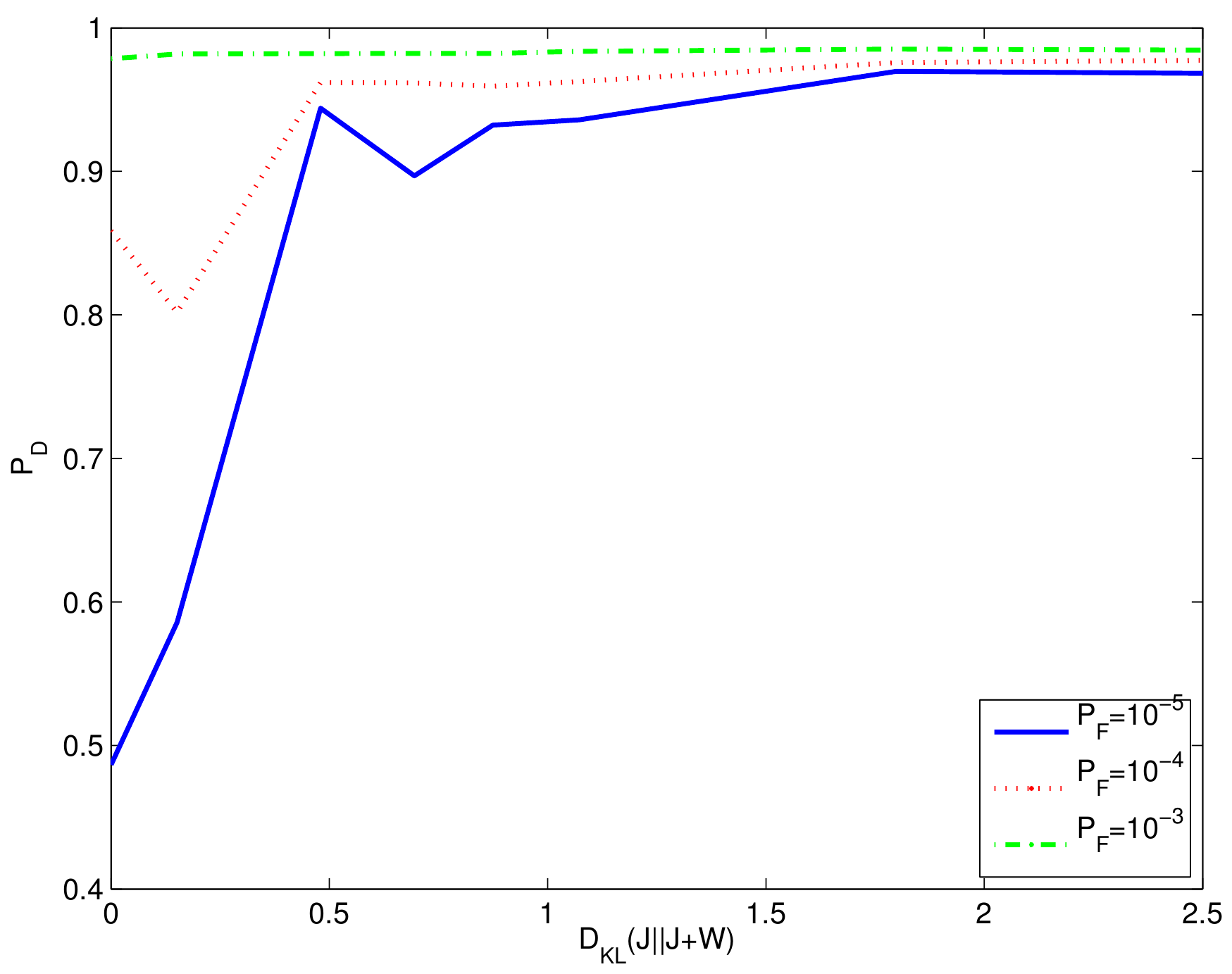}
  \caption{$P_D$ vs the detectability for fixed $P_F$ in Scenario A with $L=5$.}
  \label{fig:awSc1}
\end{figure}
\begin{figure}
  \centering
	    \includegraphics[width=0.9\columnwidth]{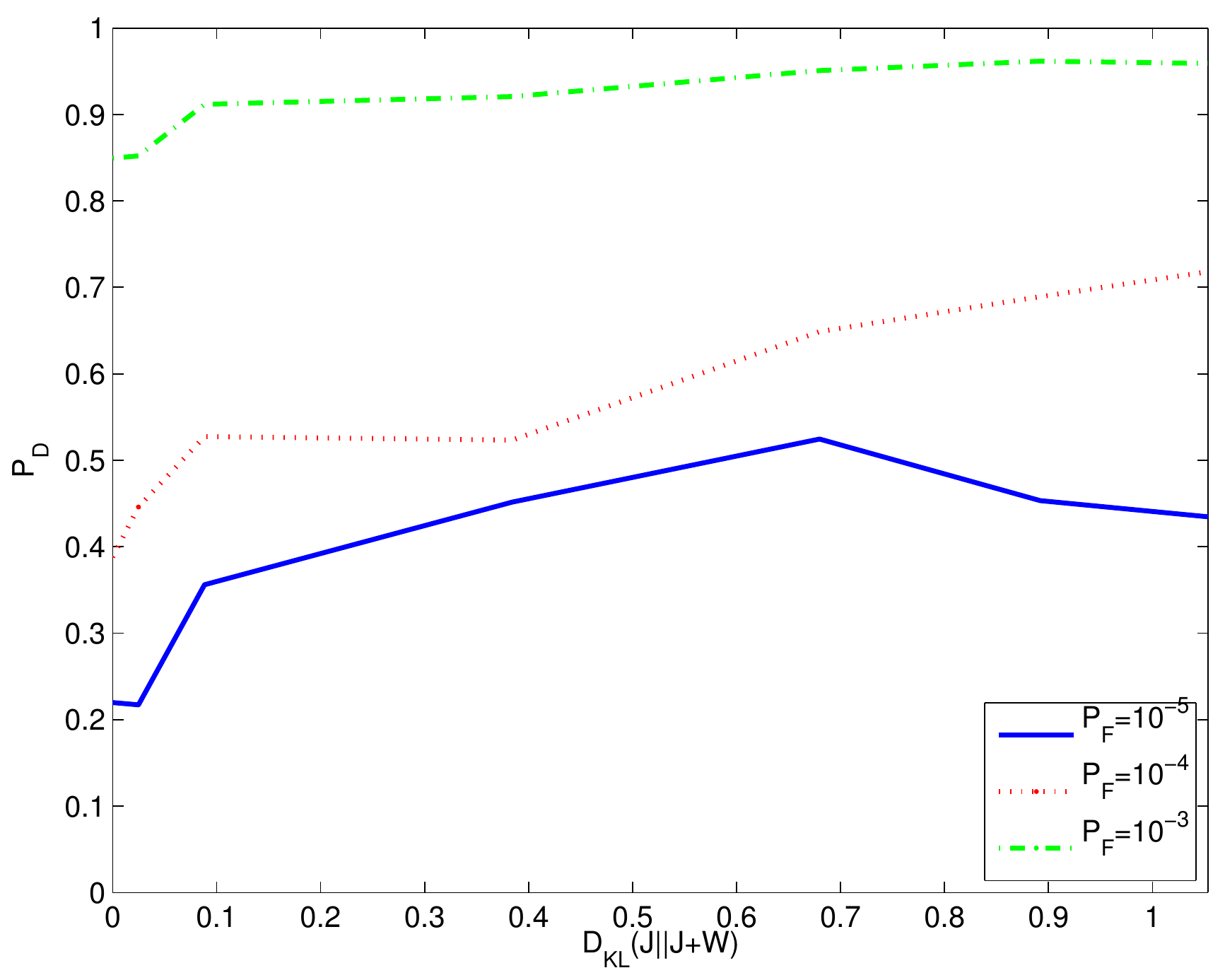}
	\caption{$P_D$ vs the detectability for fixed $P_F$ in Scenario B with $L=25$.}
  \label{fig:awSc2}
\end{figure}

We simulate Scenario A with $L=5$ and Scenario B with $L=25$ under no traffic modification, where we evaluate the trade-off between the detectability and $P_D$ when $P_F$ is fixed. Results are depicted in Figures \ref{fig:awSc1} and \ref{fig:awSc2}, where we can see that watermarking schemes give a significant improvement under low-jitter conditions even with $W_{max}=2$ ms, (cf. $D_{KL}(J||J+W)=0.486$), but this improvement is significantly lower on large-jitter conditions, e.g. the Tor network, even of very large $W_{max}$ , for instance, for $W_{max}=250$ ms (cf.$D_{KL}(J||J+W)=0.679$).

\section{Comparison with other schemes}\label{sc:com}
We want to compare our passive analysis with four other state-of-the-art traffic watermarking schemes: IB~\cite{PyPaReWaNi:12}, ICB~\cite{WaChJa:07} , RAINBOW~\cite{HoKiBo2:09} and SWIRL~\cite{HoKi:11}.To this end, we extend our simulator to be able to embed the mentioned watermarks and to detect them. 

The presented results have been obtained with the following parameters: IB, ICB and SWIRL use a time interval of 500 ms; this is the value used in the original ICB experiments reported in~\cite{WaChJa:07}.The experiments for SWIRL in~\cite{HoKi:11} use 2 s, but with short sequences this implies that many flows cannot be watermarked as the whole flow falls into one interval. We compensate this shorter interval by dividing it into less subintervals (5 instead of 20). In our experiments RAINBOW can modify the IPD up to 20 ms, which is the largest watermark amplitude used in the simulations in~\cite{HoKiBo2:09}.

\begin{figure}
  \centering
    \includegraphics[width=0.9\columnwidth]{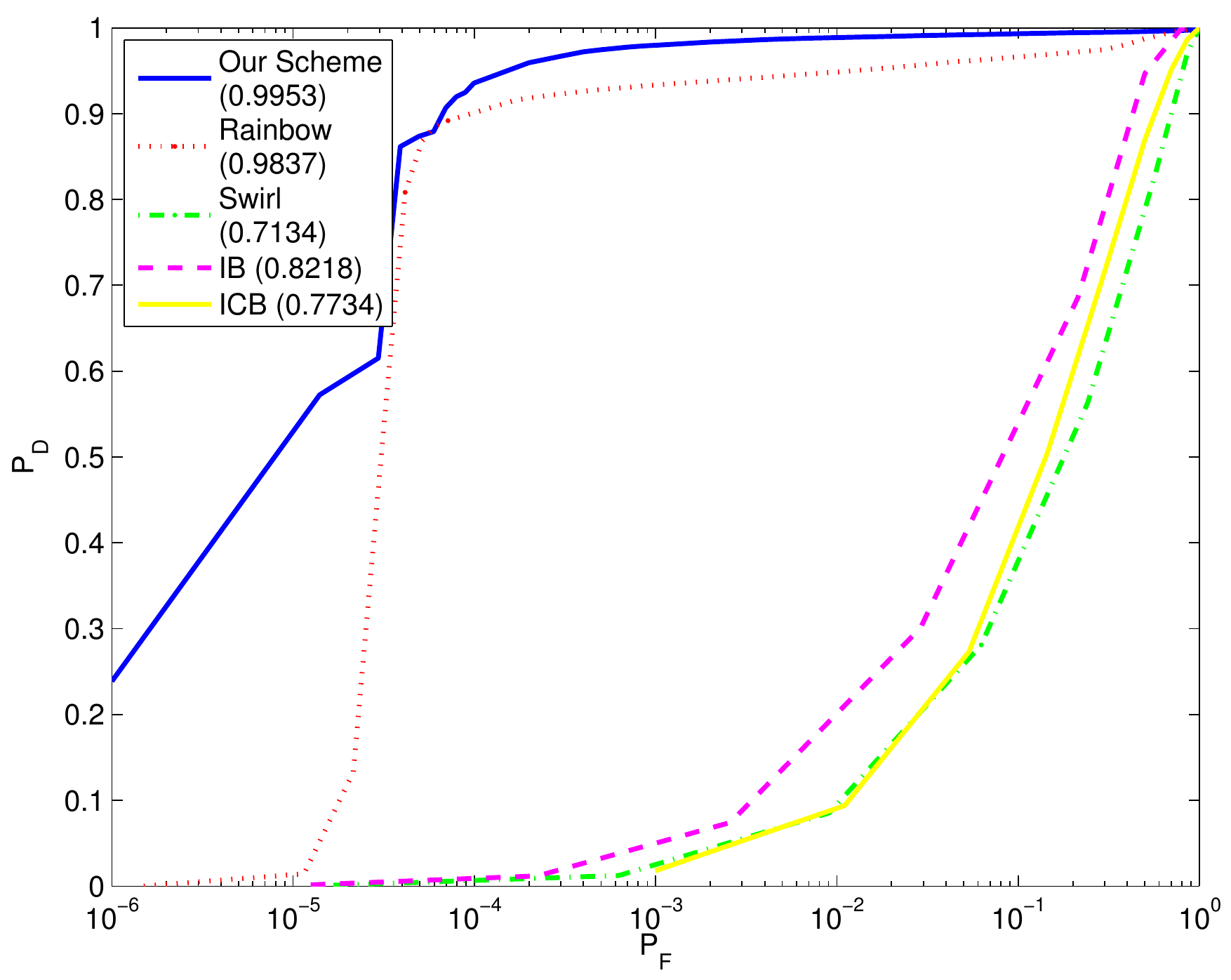}
  \caption{Comparison of algorithms on Scenario A with $L=5$.}
  \label{fig:c1Sc1}
\end{figure}

\begin{figure}
  \centering
   \includegraphics[width=0.9\columnwidth]{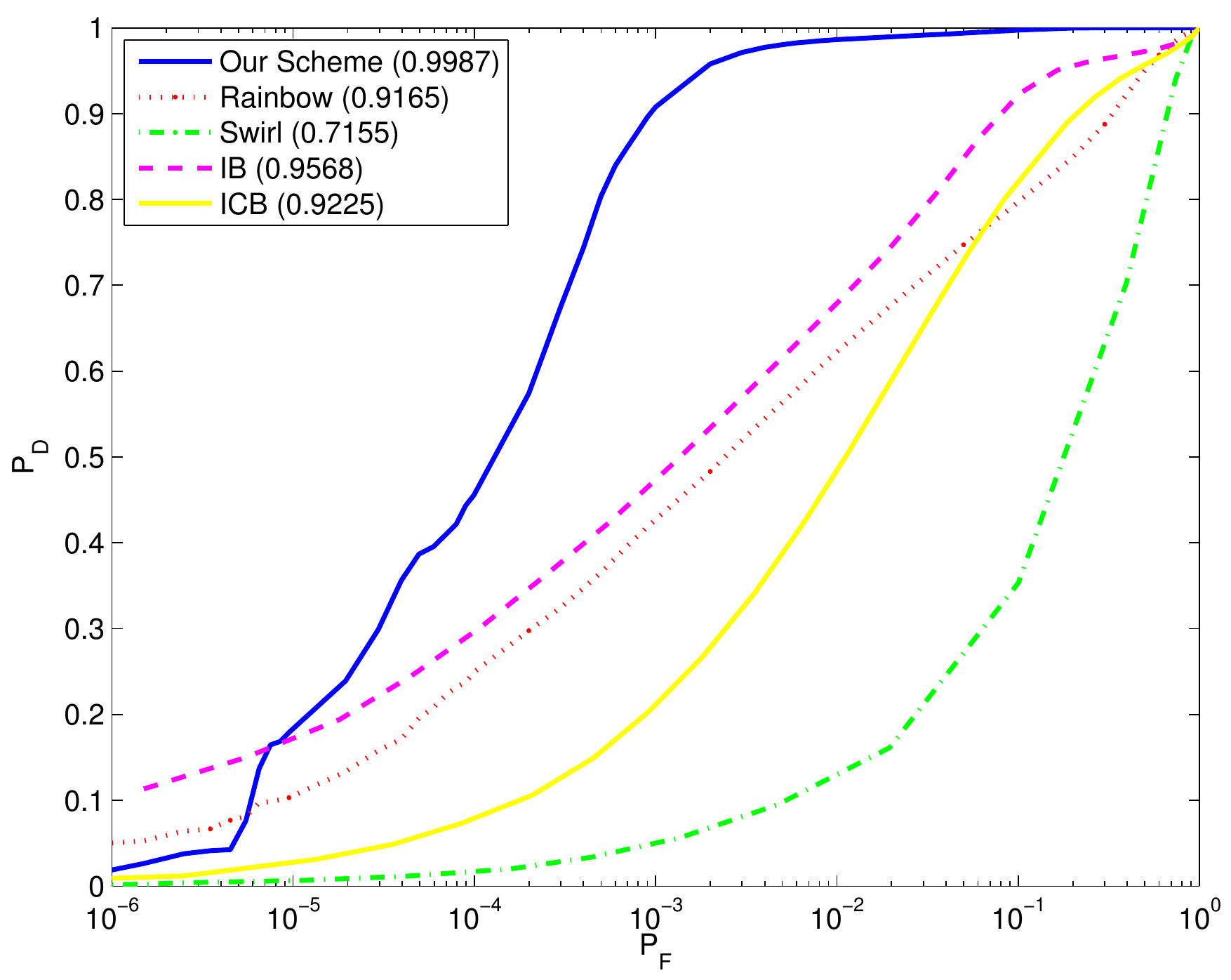}
  \caption{Comparison of algorithms on Scenario B with $L=50$.}
  \label{fig:c1Sc2}
\end{figure}

We first compare the performance in both scenarios when the flows do not suffer any addition or removal of packets, for this we use \eqref{eq:test1C}.  We take $L=5$ in Scenario A and $L=50$ in Scenario B. Figure \ref{fig:c1Sc1} shows the results for Scenario A, where  our scheme and RAINBOW outperform the rest by a significant amount. This is due to the fact that both are non-blind and perform good with short sequences if the PDV has small variance. The  other watermarking schemes do not perform well with short sequences. Figure \ref{fig:c1Sc2} shows the results in Scenario B. We see that with longer sequences IB and ICB despite of the larger PDV sequence improve their performance.


\begin{figure}
  \centering
    \includegraphics[width=0.9\columnwidth]{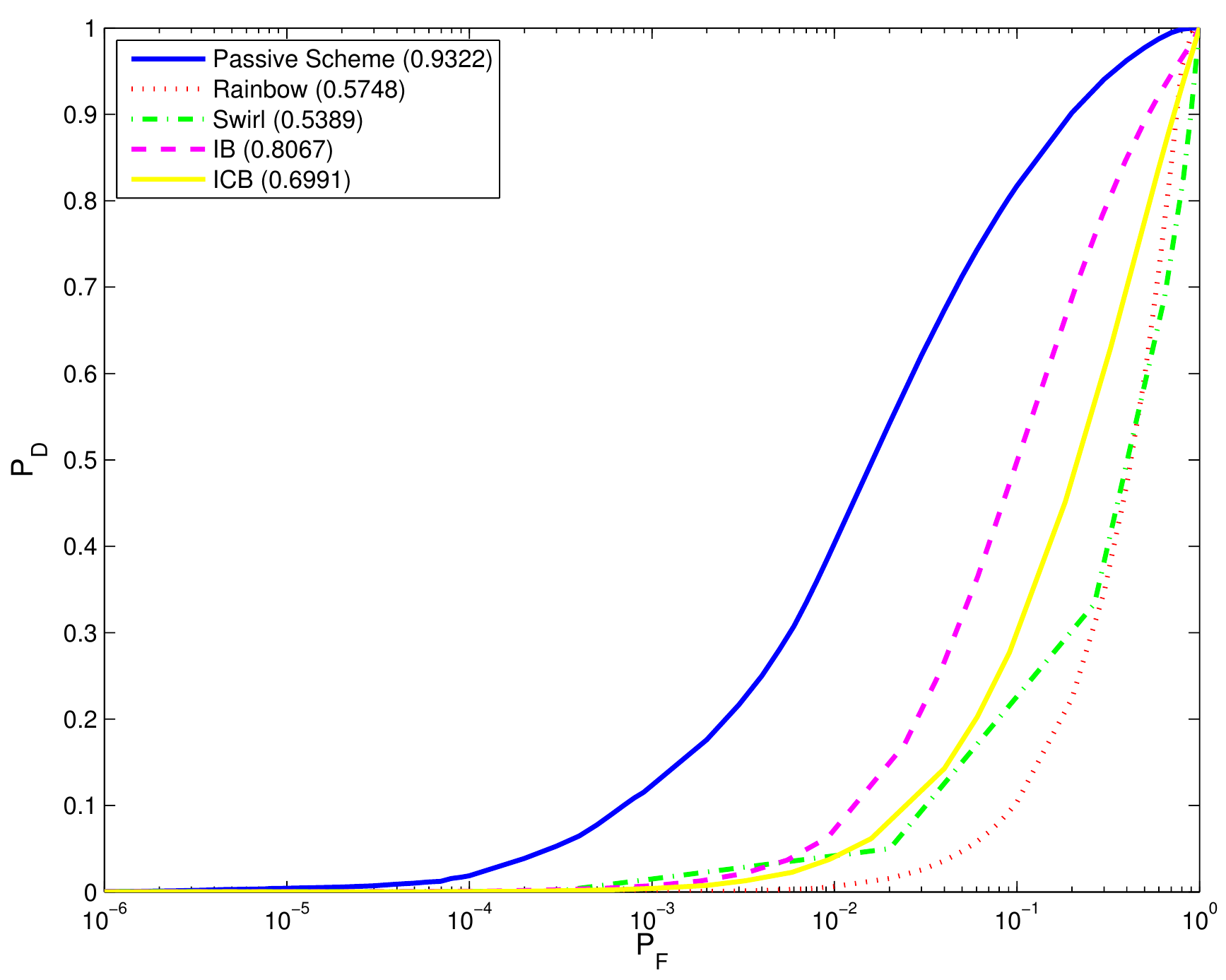}
  \caption{Comparison of algorithms on Scenario A under flow modification with $ L=50$.}
  \label{fig:c2Sc1}
\end{figure}

\begin{figure}
  \centering
    \includegraphics[width=0.9\columnwidth]{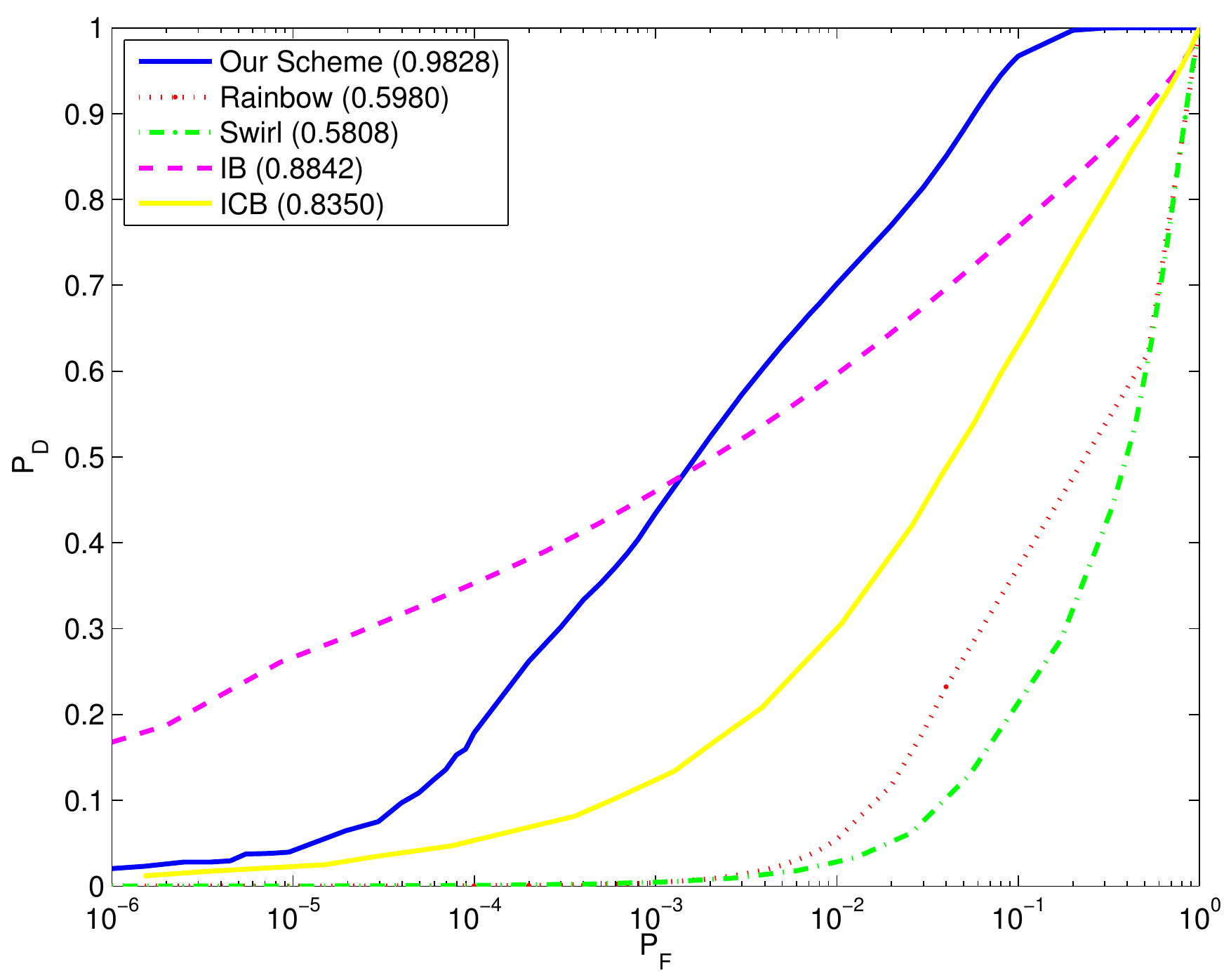}
  \caption{Comparison of algorithms on Scenario B under flow modification with $ L=250$.}
  \label{fig:c2Sc2}
\end{figure}

We also compare the performance under traffic modification using Attack 5a, i.e., 500\% chaff traffic added, $S=2$ and random delays with $A_{max}=50$ ms. As before, we fix $L=50$ in Scenario A and $L=250$ in Scenario B. Results are shown in Figures \ref{fig:c2Sc1} and \ref{fig:c2Sc2}. Note that RAINBOW or SWIRL are not designed to be robust against an active attacker.

Our algorithm is more robust to the considered traffic modifications than the rest of schemes, for example, in Scenario B, we achieve $AUC=0.9828$, while IB achieves $AUC=0.8842$, ICB $AUC=0.8350$, and for both RAINBOW and SWIRL $AUC<0.6$. Recall also that we do not modify the flow, while the rest do.

Our scheme performs better than RAINBOW, which is also a non-blind detection, although it does not modify the IPDs. The improvement in performance is due to using a likelihood test (optimal) instead of normalized correlation. Recall also that the IPDs have been restricted to be larger than 10 ms. Lifting this restriction would have a bigger impact on passive analysis than on a watermarking scheme.

\section{Real Implementation}\label{sc:ri}
Obviously, simulations are not fully realistic. To check if simulator results are applicable to real networks, we carry out a real implementation of the proposed passive analysis scheme, the watermark modification proposed in Section \ref{sc:watvspa} and the watermark schemes with which we compared in Section \ref{sc:com} for A and B scenarios.

For the first experiment, we launched three EC2~\cite{AWS} instances. We used replayed SSH connections from real traces taken at University of Vigo and the stepping stone was created by forwarding the traffic with the \texttt{socat} command. For the second experiment, we replay connections from real HTTP traces also from University of Vigo. We use 6 packets ($L=5$) and 51 ($L=50$) for Scenarios A and B, respectively. The experiment is repeated 1000 times in each case. In order to obtain values of the test under $H_0$, we use the saved timing information from the previous sequence in the non-blind cases, i.e., our proposed method and RAINBOW, and for the blind cases, i.e., IB, ICB and SWIRL, we use a different random key. 

The parameters chosen are maximum IPD variation for RAINBOW and watermark modification of 5 ms in Scenario A and 20 ms in Scenario B, that are the middle and maximum amplitudes in the experiments presented in~\cite{HoKiBo2:09}. For the blind-watermark, IB and ICB uses a interval size of 500 ms, SWIRL uses an interval length of 250 ms and 1000 ms for Scenarios A and B, respectively, divided into 5 subintervals of 3 slots each. These values have been chosen to maximize the AUC in each scenario.

Experiments are carried out in a non-active-attack scenario, this means that insertions and losses are only due to repacketization. As the detector from\eqref{eq:test2} needs a value for $P_L$, we use $P_{NL}$ from Table \ref{tab:delst}.

\begin{figure}
  \centering
    \includegraphics[width=0.9\columnwidth]{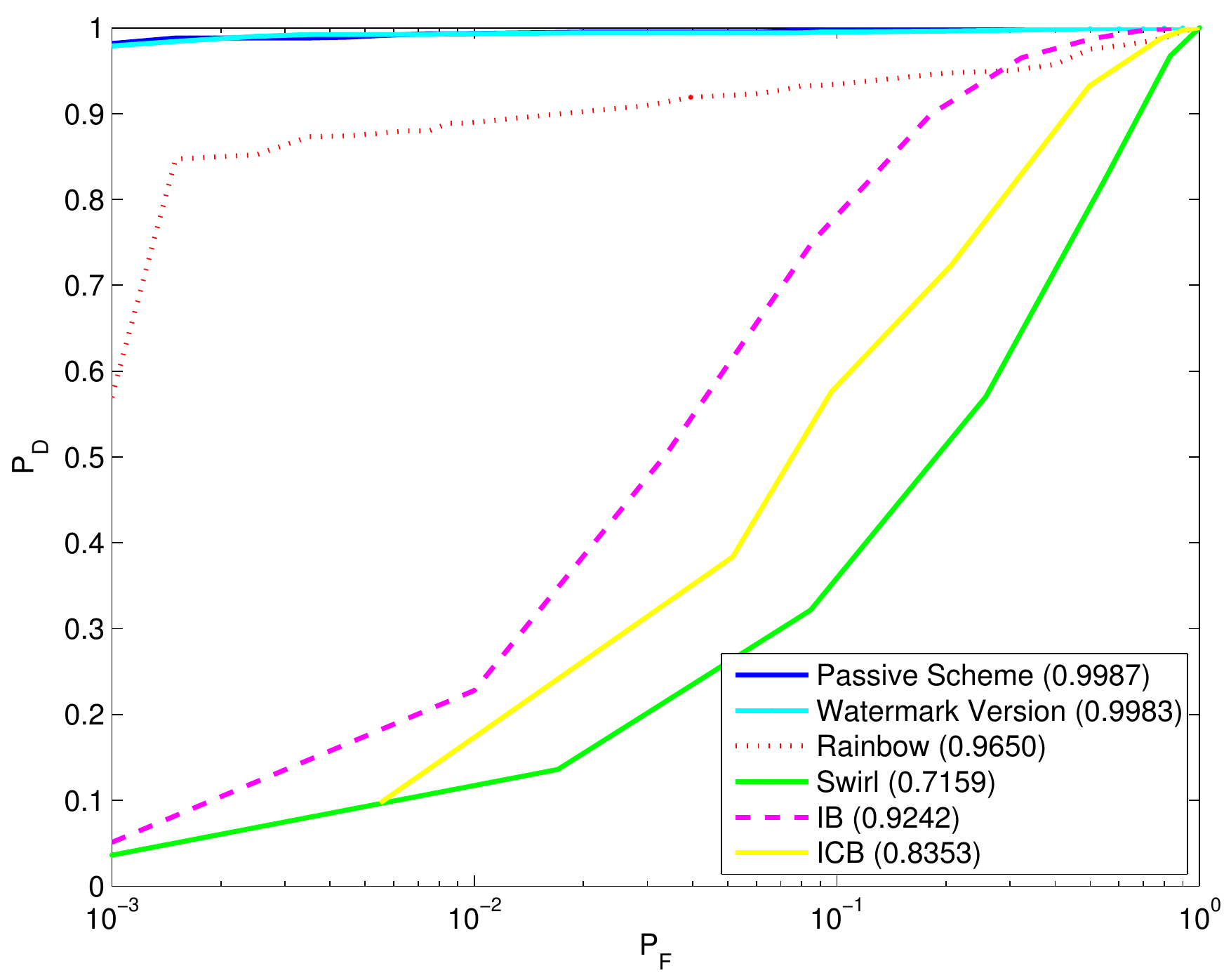}
  \caption{Real Implementation on Scenario A with $L=5$.}
  \label{fig:rSc1}
\end{figure}

\begin{figure}
  \centering
    \includegraphics[width=0.9\columnwidth]{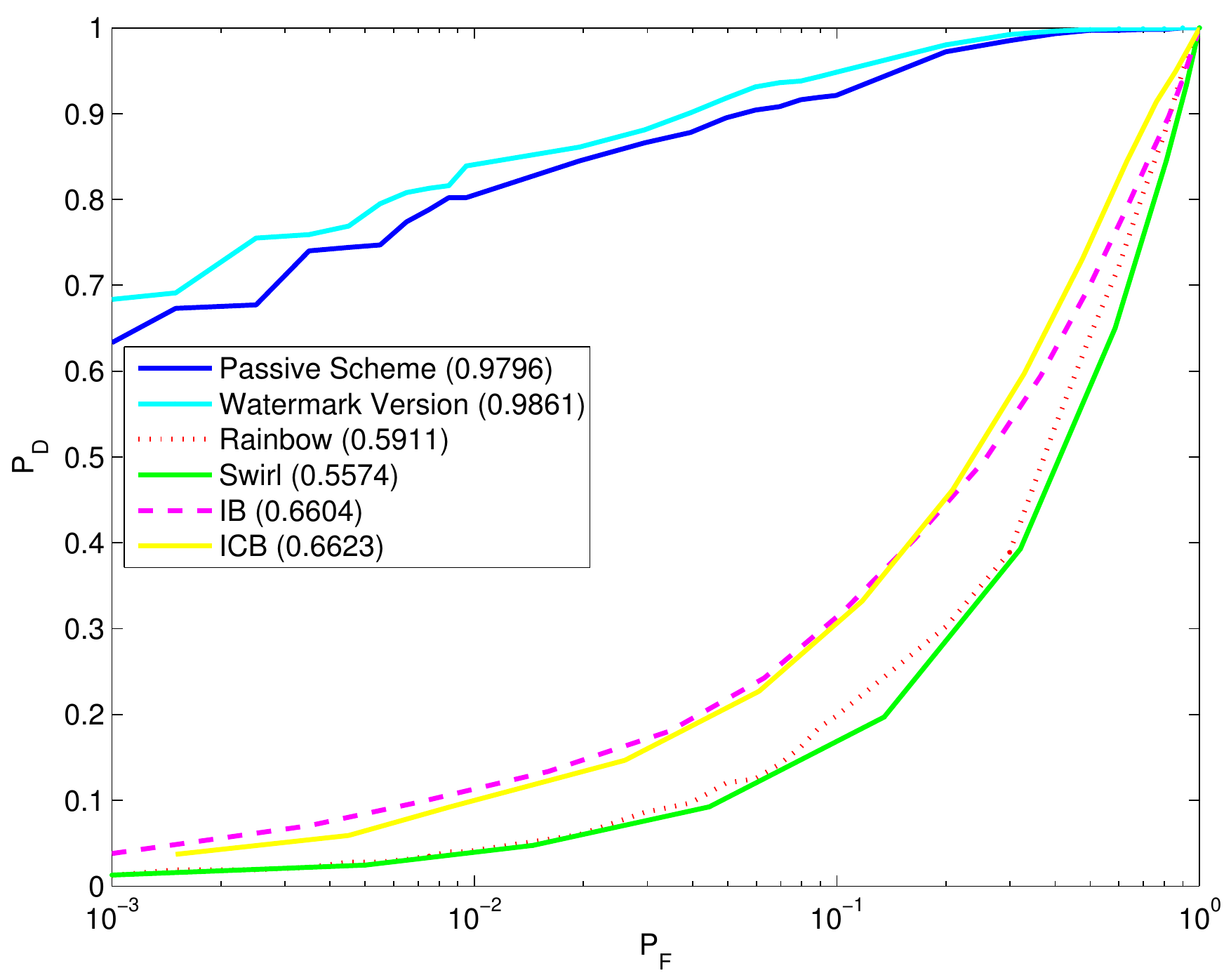}
  \caption{Real Implementation on Scenario B with $L=50$.}
  \label{fig:rSc2}
\end{figure}

Results in Scenario A are similar to the simulator results: $AUC=0.9987$ for Real Scenario vs $AUC=0.9983$ for the Simulator. However, Scenario B shows a decrease in performance for the Real Scenario compared to the simulator results. This loss of performance affects all schemes, being for ours less severe.

\section{Conclusions}\label{sc:con}

Network flow watermarks are becoming increasingly popular in traffic analysis owing to their improved performance as compared to passive analysis. Unfortunately, the ease with which these watermarks can be exposed has revealed itself as the Achilles' heel of these techniques and can lead to a traffic modification attack in which the watermark is finally removed. In this paper we have presented a highly-optimized traffic analysis method for deciding if two flows are linked that can be used as passive analysis, as well as a watermarking scheme.

With performance in mind, we develop an optimal decoder, i.e. likelihood-ratio test, that allows to achieve a very good performance under a passive analysis scheme. For example, with 21 packets separated at least 10 ms we can correlate two flows obtaining $P_D = 0.9861$ given a false  alarm probability equal to $10^{-5}$ without flow modifications.

A more robust detector is created that can deal with chaff traffic, flow splitting and random delays added by an attacker. To this end, packet matching is carried out by removing the packets that do not have a correspondent in the other flow. Then, a new likelihood-ratio test that considers losses and the maximum delay that an attacker can add is derived.

Afterwards, we study the trade-off between performance improvement versus the detectability on a watermarking scheme based on our algorithm. We also show a comparison with four state-of-the-art traffic watermarking schemes. Finally, a real implementation is carried out to show that the simulator results can be extended to real networks. 

The obtained results show that passive analysis schemes with an optimal detector can compete with and outperform state-of-the-art traffic watermarking schemes, giving the advantages
of being undetectable, which decreases the risk of a traffic modification attack, and that they can be carried out ex-post, in addition to in real-time, allowing them to be used in forensic analysis applications as well as in intrusion detection. 

\section*{Acknowledgment}
The authors would like to thank Dr. Negar Kiyavash for her insightful and helpful comments. This work was supported in part by Iberdrola Foundation through the Prince of Asturias Endowed Chair in Information Science and Related Technologies.

\bibliographystyle{IEEEtran}
\bibliography{IEEEabrv,../../bib/watermark,../../bib/traces,../../bib/gametheory}

\end{document}